\documentclass[prd,preprintnumbers,nofootinbib]{revtex4} 
\usepackage{graphicx} 
\usepackage{amsmath}
\usepackage{amsfonts,amsbsy}
\usepackage{amssymb}
\usepackage{appendix}
\usepackage{array}
\usepackage{multirow}
\def\be{\begin{equation}}
\def\ee{\end{equation}}
\def\bea{\begin{eqnarray}}
\def\eea{\end{eqnarray}}

\def\eq#1{{Eq.~(\ref{#1})}}
\def\fig#1{{Fig.~\ref{#1}}}

\def\beq{\begin{equation}}
\def\eeq{\end{equation}}
\def\bea{\begin{eqnarray}}
\def\eea{\end{eqnarray}}
\newcommand{\Lb}{\left(}
\newcommand{\Rb}{\right)}

\newlength{\dinwidth}
\newlength{\dinmargin}
\renewcommand{\vec}[1]{\boldsymbol{#1}}
\newcommand{\dif}{\mathrm{d}}

\makeatother

\makeatother

\begin{document}

%\title{ Impact-parameter dependent saturation model and new combined HERA data }
%\title{ Dips in Diffractive Vector Meson production at the LHC and gluon saturation}
\title{ Exclusive vector meson production at high energies and gluon saturation}
\author{N\'estor Armesto$^{1}$ and Amir H. Rezaeian$^{2,3}$}
%\author{Amir H. Rezaeian$^{1,2}$ et al.}
\affiliation{\it ~$^1$ Departamento de F\'isica de Part\'iculas and IGFAE, Universidade de Santiago de Compostela,
15782 Santiago de Compostela, Galicia, Spain \\
%~$^2$ PH Department, TH Unit, CERN, CH-1211 Geneva 23, Switzerland \\
$^2$ Departamento de F\'\i sica, Universidad T\'ecnica
Federico Santa Mar\'\i a, Avda. Espa\~na 1680,
Casilla 110-V, Valparaiso, Chile\\
$^3$  Centro Cient\'\i fico Tecnol\'ogico de Valpara\'\i so (CCTVal), Universidad T\'ecnica
Federico Santa Mar\'\i a, Casilla 110-V, Valpara\'\i so, Chile}
\begin{abstract}
We systematically study exclusive diffractive (photo) production of vector mesons ($J/\psi$, $\psi(2s)$, $\phi$ and $\rho$) off protons in high-energy collisions and investigate whether the production is a sensitive probe of gluon saturation. We confront saturation-based results for diffractive $\psi(2s)$ and $\rho$  production at HERA and $J/\psi$ photoproduction with all available data including recent ones from HERA, ALICE and LHCb, finding good agreement.  In particular, we show that the $t$-distribution of differential cross sections of photoproduction of vector mesons offers a unique opportunity to  discriminate among saturation and non-saturation models. This is due to the emergence of a pronounced dip  (or multiple dips) in the $t$-distribution of diffractive photoproduction of vector mesons at relatively large, but potentially accessible $|t|$ that can be traced back to the unitarity features of colour dipole amplitude in the saturation regime. We show that in saturation models the dips in $t$-distribution recede towards lower $|t|$  with decreasing mass of the vector meson, increasing energy or decreasing  Bjorken-$x$, and decreasing virtuality $Q$.  We provide  various predictions for exclusive (photo) production of different vector mesons including the ratio of $\psi(2s)/J/\psi$ at HERA, the LHC, and future colliders. 
  
\end{abstract}
\maketitle

\section{Introduction}
There is strong theoretical evidence that quantum chromodynamics (QCD) at high-energy (or small Bjorken-$x$) leads to a non-linear regime where gluon recombination or unitarity effects become important \cite{sg,mv}, resulting in a saturation of parton densities in hadrons and nuclei.
The quest for experimental evidence of the possible signature of gluon saturation phenomenon has been the  program of various past or existing experiments from HERA and RHIC to the LHC, and future experiments such as an Electron-Ion Collider (EIC) \cite{eic} and the LHeC \cite{lhec}.  Nevertheless, experimental evidence that can unarguably point towards gluon saturation phenomenon, has been elusive so far. This is because the experiments currently at our disposal are limited in their kinematic coverage, and often other approaches provide alternative descriptions of the same sets of data. 

An effective field theory approach that describes the high-energy limit of QCD is the colour glass condensate (CGC), see the review \cite{Gelis:2010nm}. In this formalism, the standard quantum evolution equations (with large logarithms of $1/x$ resummed), lead to a situation in which the occupancy of the slow  modes in the hadron is so high than they can be treated classically, with the fast modes considered as sources. 
The corresponding renormalisation group equations, known in the limit of scattering of a dilute probe on a dense hadron, are the so-called Jalilian-Marian-Iancu-McLerran-Weigert-Leonidov-Kovner (JIMWLK) hierarchy of equations \cite{jimwlk} or, in the large $N_c$ limit, the Balitsky-Kovchegov (BK) equation \cite{bk}, presently known to next-to-leading accuracy \cite{Balitsky:2008zza,Kovner:2013ona}.
%\cite{bk,bb}.

One of the most crucial tests of the CGC (or saturation) approach has been its success in the description of the highly precise combined data of the proton structure at HERA  \cite{rcbk,ip-sat-a,b-cgc-a} alongside data from exclusive diffractive processes in electron-proton collisions, such as exclusive vector meson production and deeply virtual Compton scattering (DVCS) \cite{ip-sat-a,b-cgc-a}.  Nevertheless, the standard DGLAP-type approaches - without inclusion of  any saturation effect - give an equally good description of the same data. While the CGC description can be considered more economical due to the use of a significantly smaller number of fitting parameters, it is limited to small-$x$ data and restricted to the gluon sector. On the other hand, in addition to DIS and diffractive processes \cite{rcbk,ip-sat-a,b-cgc-a,ip-sat0, ip-sat1, watt-bcgc,ip-g2}, within the CGC framework it is also possible to simultaneously describe other high-energy hadronic interactions in a regime not currently accessible to approaches that rely on collinear factorisation. For example, in proton-proton \cite{pp-LR} and nuclear collisions \cite{jav-d,aa-LR,hic-ap, raju-glasma,all-pa} several observables have been successfully addressed: single inclusive hadron \cite{pa-raju,jav-pa,pa-R,pa-t,pa-jam,pa-ana} and prompt photon \cite{pa-R,jr-p} production, and semi-inclusive photon-hadron \cite{jr-p,p-h} and dihadron \cite{di-all} productions. For a recent review, see Ref.\,\cite{Albacete:2014fwa} and references therein.

Exclusive diffractive vector meson production provides a rich testing ground of many QCD novel properties \cite{ip-sat-a,b-cgc-a,ip-sat0, ip-sat1, watt-bcgc,ip-g2,diff3,ann2,diff4,stan,diff2,ggg,exc}. In particular, by measuring the squared momentum transfer $t$, one can study the transverse spatial distribution of the gluons in the hadron wave function that cannot be probed in inclusive DIS. In this respect, new experimental measurements are under way. The LHCb and ALICE collaborations have recently released new data on $J/\psi$ photoproduction with  photon-proton center-of-mass energies up to about 1.3 TeV \cite{lhcb,lhcbn,TheALICE:2014dwa}, the highest energy ever measured so far in this kind of reaction. 
Alongside this, the H1 Collaboration also recently reported some new data for  $J/\psi$ with improved precision \cite{h1-2013}.  On the other hand, the recently released high-precision combined HERA data \cite{Aaron:2009aa,Abramowicz:1900rp}  that were not available at the time of previous studies of  diffractive processes \cite{eic,lhec,ip-sat0, ip-sat1, watt-bcgc,watt-2}, provide extra important constraints on saturation models \cite{ip-sat-a,b-cgc-a}.

In this work, we analyse these data on exclusive photoproduction of vector mesons off the proton and provide predictions for the kinematics accessible in future experiments. We show that the freedom to choose the charm mass in the range consistent with global analysis of inclusive observables, results in sizable uncertainties for the total cross-section of elastic photoproduction of vector mesons.  Nevertheless, we show that, even with these uncertainties,  the recent LHC data \cite{lhcbn,TheALICE:2014dwa} seem to favour the saturation picture. We systematically study elastic diffractive production of different vector mesons $J/\psi$, $\psi(2s)$, $\phi$ and $\rho$  off protons and  investigate which vector meson production is more sensitive to saturation physics and what measurement can potentially be  a better probe of the signal.  In particular, we study $\psi(2s)$ diffractive production by constructing the $\psi(2s)$ forward wave function via a fit to the leptonic decay, and we provide various predictions for diffractive $\psi(2s)$ production as well as the ratio of $\psi(2s)/J/\psi$ at HERA and the LHC. 
Furthermore,  we find that the corresponding $t$-distributions of differential cross-section may unambiguously  discriminate among saturation and non-saturation models. This is due to the emergence of a pronounced dip  (or multiple dips) in the $t$-distribution of diffractive photoproduction of vector mesons at relatively large $|t|$ (but within reach of future experiments \cite{eic,lhec}) which is directly related to saturation physics.  In this way, we go beyond existing recent works on $J/\psi$ and $\psi(2s)$ production both in the dipole model \cite{Ducati:2013tva,Goncalves:2014wna} and in pQCD \cite{n-pqcd,Jones:2013eda,Guzey:2014axa,Guzey:2013qza}, and of lighter mesons in the dipole model \cite{Santos:2014vwa}.

This paper  is organised as follows. In section II, we introduce the  formulation of the colour dipole approach for calculating  exclusive diffractive processes. In section III, we introduce the IP-Sat and b-CGC dipole models. In section IV, we present a detailed numerical analysis and our main results.  In subsection A we first show our results and predictions for the total diffractive cross-section of different vector mesons, while in subsection B  we discuss the origin of the dips in the $t$-distribution of diffractive photoproduction of vector mesons and  provide predictions for future experiments. We summarise our main results in section V.

%%%%%%%%%%%%%%%%%%%%%%%%%%%%%%%%%%%
%%%%%%%%%%%%%%%%%%%%%%%%%%%%%%%%%%%
\section{ Exclusive diffractive processes in the colour-dipole formalism}
In the colour dipole formalism, the underlying mechanism for diffractive production of different vector mesons and for inclusive DIS is similar. Namely, one must calculate the probability of finding a colour dipole of transverse size $r$ with impact parameter $b$ in the wave function of a (real or virtual) photon or of a vector meson. Similar to the case of the inclusive DIS process,  the scattering amplitude for the exclusive diffractive process $\gamma^*+p\to V+p$, with a final state vector meson $V=J/\psi, \psi(2s), \phi,\rho$  (or a real photon $V=\gamma$  in DVCS), can be written in terms of a convolution of the dipole amplitude $\mathcal{N}$ and the overlap of the wave functions of the photon and the exclusive final state particle (see \cite{ip-sat1,ip-sat-a, b-cgc-a} and the references therein), 
\begin{equation} \label{am-i}
  \mathcal{A}^{\gamma^* p\rightarrow Vp}_{T,L} = \mathrm{2i}\,\int\!\dif^2\vec{r}\int\!\dif^2\vec{b}\int_0^1\!\dif{z}\;(\Psi_{V}^{*}\Psi)_{T,L}(r,z,m_f,M_V;Q^2)\;\mathrm{e}^{-\mathrm{i}[\vec{b}-(1-z)\vec{r}]\cdot\vec{\Delta}}\mathcal{N}\left(x,r,b\right),
\end{equation}
with $\vec{\Delta}^2=-t$ and $t$ being the squared momentum transfer. In this equation, $\mathcal{N}$ is the imaginary part of the forward $q\bar{q}$ dipole-proton scattering amplitude with transverse dipole size $r$ and impact parameter $b$. The parameter $z$ is the fraction of the light cone momentum of the virtual photon carried by the quark and $m_f$ denotes the mass of the quark with flavour $f$. 
The above expression can be understood in light front time as distinct chronological subprocesses, namely the $\gamma^\star$  first fluctuates into a quark-antiquark pair (the so-called colour $q\bar{q}$-dipole)  which then interacts with the target. Finally the $q\bar{q}$ pair recombines to form the final state vector meson. In \eq{am-i} summations over the quark helicities and over the quark flavour $f=u,d,s, c$ are implicit. The phase factor $\exp\left(i(1-z)\vec{r}\cdot\vec{\Delta}\right)$ in the above equation is due to the non-forward wave-function contribution \cite{bbb}. In \eq{am-i}, the $\Psi_{V}^{*}\Psi$ is the forward overlap wave function of photon and vector meson (see below).

%%%%%%%%%%%%%%%%%%%%%%%%%%%%%%%%%%%

The differential cross-section of the exclusive diffractive processes can then be written in terms of the scattering amplitude as \cite{ip-sat1,watt-bcgc,b-cgc-a}, 
\begin{equation}
  \frac{\dif\sigma^{\gamma^* p\rightarrow Vp}_{T,L}}{\dif t} = \frac{1}{16\pi}\left\lvert\mathcal{A}^{\gamma^* p\rightarrow Vp}_{T,L}\right\rvert^2\;(1+\beta^2) R_g^{2},
  \label{vm}
\end{equation}
with
\bea \label{eq:beta} 
  \beta &=& \tan\left(\frac{\pi\delta}{2}\right), \nonumber\\
R_g(\delta) &=& \frac{2^{2\delta+3}}{\sqrt{\pi}}\frac{\Gamma(\delta+5/2)}{\Gamma(\delta+4)}, \nonumber\\
\delta &\equiv& \frac{\partial\ln\left(\mathcal{A}_{T,L}^{\gamma^* p\rightarrow Vp}\right)}{\partial\ln(1/x)}, \
\eea
where the factor $(1+\beta^2)$ takes into account the missing real part of amplitude (notice that the amplitude in \eq{am-i} is purely imaginary), with $\beta$ being the ratio of real to imaginary parts of the scattering amplitude \cite{ip-sat1,ip-sat-a,b-cgc-a}.  The factor $R_g$ incorporates the skewness effect, coming from the fact that the gluons attached to the $q\bar{q}$ can carry different light-front fractions $x,x^{\prime}$ of the proton \cite{ske,mrt,ske-n}.
The skewedness factor given in \eq{eq:beta}  was obtained at NLO level, in the limit that $x^{\prime}\ll x\ll 1$ and at small $t$ assuming that the diagonal gluon density of target has a power-law form  \cite{ske}.  Note that there are uncertainties with respect to the actual incorporation of the skewness correction at small $x$ in dipole models\footnote{In the IP-Sat model \cite{ip-sat-a}, the skewness effect can be simply incorporated by  multiplying the gluon distribution $xg(x,\mu^2)$ by a factor $R_g(\gamma)$ with $\gamma \equiv \frac{\partial\ln\left[xg(x,\mu^2)\right]}{\partial\ln(1/x)}$. This is consistent with the prescription given in Eqs.\,(\ref{vm}, \ref{eq:beta}) \cite{b-cgc-a}.} \cite{Rezaeian:2013eia}. However, these uncertainties will not affect our main results and conclusions.

%%%%%%%%%%%%%%%%%%%%%%%
%%%% \section{Forward wave functions for $J/\psi$ and $\psi(2s)$}
The forward photon wave functions at leading order is well known in QCD, see e.g. Refs.~\cite{Dosch,Lepage}.  The normalized photon wave function for the longitudinal photon polarization ($\lambda = 0$) and the transverse photon polarisations ($\lambda = \pm 1$) are given by~\cite{beta1}, 
\bea
  \Psi_{h\bar{h},\lambda=0}(r,z,Q) &=&  e_f \sqrt{4\pi\alpha_{\mathrm{em}}}\, \sqrt{N_c}\, 
  \delta_{h,-\bar h} \, 2Qz(1-z)\, \frac{K_0(\epsilon r)}{2\pi}, \\
  \Psi_{h\bar{h},\lambda=\pm 1}(r,z,Q) &=&
  \pm e_f \sqrt{4\pi\alpha_{\mathrm{em}}} \, \sqrt{2N_c}\,
  \left\{
  \mathrm{i}e^{\pm \mathrm{i}\theta_r}[
    z\delta_{h,\pm}\delta_{\bar h,\mp} - 
    (1-z)\delta_{h,\mp}\delta_{\bar h,\pm}] \partial_r \, + \, 
  m_f \delta_{h,\pm}\delta_{\bar h,\pm}
  \right\}\, \frac{K_0(\epsilon r)}{2\pi},
  \label{tspinphot}
\eea
where $N_c$ is the number of colours,  the subscripts $h$ and $\bar h$ denote the helicities of the quark and the antiquark respectively and $\theta_r$ is the azimuthal angle between the vector $\vec{r}$ and the $x$-axis in the transverse plane. We have used a notation $\epsilon^2 \equiv z(1-z)Q^2+m_f^2$ where  the subscript $f$ denotes the flavour. 

Following Refs.\,\cite{ip-sat0,ip-sat1,ip-sat-a,b-cgc-a,beta,FF}, we assume that the forward vector meson wave functions $\Psi_V$ are effectively dominated by the $q\bar{q}$ Fock component and have the same spin and polarization structure as in the case of the  photon:
\bea
  \Psi^V_{h\bar{h},\lambda=\pm 1}(r,z) &=&
  \pm\sqrt{2N_c}\, \frac{1}{z(1-z)} 
  \left\{
  \mathrm{i}e^{\pm \mathrm{i}\theta_r}[
    z\delta_{h,\pm}\delta_{\bar h,\mp} - 
    (1-z)\delta_{h,\mp}\delta_{\bar h,\pm}] \partial_r \, + \, 
  m_f \delta_{h,\pm}\delta_{\bar h,\pm}
  \right\}\, \phi_T(r,z), \\
  \Psi^V_{h\bar{h},\lambda=0}(r,z) &=& \sqrt{N_c}
  \delta_{h,-\bar h} 
  \left[ M_V\,+ \, \delta \, \frac{m_f^2 - \nabla_r^2}{M_Vz(1-z)}\,  
    \right] \phi_L(r,z),\
\eea
where $\nabla_r^2 \equiv (1/r)\partial_r + \partial_r^2$, $M_V$ is the meson mass and the effective charge is defined $\hat{e}_f=2/3$, $1/3$, or $1/\sqrt{2}$, for $J/\psi$ (and $\psi(2s)$), $\phi$ or $\rho$ mesons respectively\footnote{See  \cite{Santos:2014vwa} for a study of the impact of different forms of the wave function on $\rho$ production.}. The longitudinally polarised vector meson wave function is slightly more complicated than in the case of the photon since the coupling of the quarks to the meson is non-local \cite{beta}.  For the scalar parts of the wave functions $\phi_{T,L}(r,z)$, we employ the boosted Gaussian wave-functions with the Brodsky-Huang-Lepage prescription  \cite{stan1}.  The boosted Gaussian wave-functions were found to provide a very good description of exclusive diffractive HERA data \cite{ip-sat1,ip-sat-a,b-cgc-a}. For the ground state vector meson ($1s$) and its first excited state $2s$,  the scalar function $\phi_{T,L}(r,z)$, has the following general form \cite{beta,FF}, 
\bea
\phi_{T,L}^{1s}(r,z) &=& \mathcal{N}_{T,L} z(1-z)
  \exp\left(-\frac{m_f^2 \mathcal{R}_{1s}^2}{8z(1-z)} - \frac{2z(1-z)r^2}{\mathcal{R}_{1s}^2} + \frac{m_f^2\mathcal{R}_{1s}^2}{2}\right), \\
  \phi_{T,L}^{2s}(r,z) &=& \mathcal{N}_{T,L} z(1-z)
  \exp\left(-\frac{m_f^2 \mathcal{R}_{2s}^2}{8z(1-z)} - \frac{2z(1-z)r^2}{\mathcal{R}_{2s}^2} + \frac{m_f^2\mathcal{R}_{2s}^2}{2}\right) \nonumber\\
&\times&\Big[1+\alpha_{2s}\left(2+\frac{m_f^2 \mathcal{R}_{2s}^2}{4z(1-z)} - \frac{4z(1-z)r^2}{\mathcal{R}_{2s}^2} - m_f^2\mathcal{R}_{2s}^2\right)\Big],\
\eea
where the parameter $\alpha_{2s}$ controls the position of the node of the radial wave function of the $V(2s)$. The boosted Gaussian wave function\footnote{ Note that the Coulomb term \cite{beta} has been ignored in the wave function here because adding it introduces another parameter to the wave function (plus an unknown running coupling) and a singular behaviour at the origin, However, this should not be important at high energy for large dipole sizes, and its contribution should be either negligible or simply absorbed into the remaining parameters of the wave function. On the phenomenology side, there is no strong evidence of Coulomb contribution even at lower energy at HERA, and indeed a good fit of vector meson  wave function to leptonic decay can be found even without it as done in \cite{b-cgc-a} and here, see table \ref{t-1}.} has several advantages over other commonly used models, namely it is more self-consistent, it is fully boost invariant and it has the proper short-distance limit at $m_f\to 0$.
 
The normalisation and orthogonality conditions allow the missing higher order Fock component of the wave functions to be effectively absorbed into the overall normalisation factor, 
\bea
 && \frac{N_c}{2\pi}\int_0^1\!\frac{\dif{z}}{z^2(1-z)^2}\int\!\dif^2\vec{r}\;
  \left\{m_f^2(\phi_T^{1s(2s)})^2+\left[z^2+(1-z)^2\right]
  \left(\partial_r\phi_T^{1s(2s)}\right)^2\right\} =1,\\ 
&&  \frac{N_c}{2\pi} \int_0^1\!
  \dif{z}\,
  \int\!\dif^2\vec{r}\;
  \left[
    M_V\phi_L^{1s(2s)}+
    \delta\,
    \frac{m_f^2-\nabla_r^2}{M_V z(1-z)}\,\phi_L^{1s(2s)}\right]^2 =1,  \\
&&\frac{N_c}{2\pi}\int_0^1\!\frac{\dif{z}}{z^2(1-z)^2}\int\!\dif^2\vec{r}\;
  \left\{m_f^2 \phi_T^{1s}\phi_T^{2s} +\left[z^2+(1-z)^2\right]
  \partial_r\phi_T^{1s} \partial_r\phi_T^{2s}\right\} =0.\
\eea
Another important input is the leptonic decay width of the vector meson which  is given by
\begin{equation}
  \Gamma_{V\to e^+e^-} = \frac{4\pi\alpha_{\rm em}^2f_V^2}{3M_V},
\end{equation}
where the decay widths are given by \cite{ip-sat0,ip-sat1}, 
\begin{gather}
  f_{V,T} = \hat{e}_f\, \left.\frac{N_c}{2\pi M_V}
  \int_0^1\!\frac{\dif{z}}{z^2(1-z)^2}
  \left\{m_f^2-\left[z^2+(1-z)^2\right]\nabla_r^2\right\}\phi_T(r,z)\right\rvert_{r=0},\\
  \label{eq:nnz_fvl}
  f_{V,L} = \hat{e}_f\, \left.\frac{N_c}{\pi}
  \int_0^1\!
  \dif{z}\,
  \left[
    M_V +\delta\, \frac{m_f^2-\nabla_r^2}{M_Vz(1-z)}\right]
  \phi_L(r,z)\right\rvert_{r=0}.
\end{gather} 
In the above, consistent with underlying dynamics of the vector meson production in the colour-dipole factorisation,  we assumed that the leptonic decay $V\to \gamma^{\star}\to e^+e^-$ can be also described by a factorized from in which the vector meson contributes mainly through its properties at the origin.  

The vector meson wave function in the  boosted Gaussian model, has only 3 (4 for $2s$ state) parameters, namely  $\mathcal{N}_{T,L}, \mathcal{R}$ and $\alpha_{2s}$ which are determined from normalisation, the orthogonality conditions and a fit to the experimental leptonic decay width. For the case of $1s$ ground state vector meson production we have $\alpha_{2s}=0$.  Unfortunately we do not have experimental data for leptonic decay width for longitudinal and transverse polarisations component separately. Therefore, we assume that the measured experimental value is the average between those for longitudinal and transverse polarisations. Note that the parameters of the wave function cannot be uniquely extracted from the conditions indicated above; namely, several sets of solutions exist.  In order to put more constrain on the parameters of the wave function, it is natural to assume that  $\mathcal{N}_{T}\approx \mathcal{N}_{L}$ (or $\mathcal{N}_{T}=\mathcal{N}_{L}$). This is because  in the boosted Gaussian wave function there is only one radius parameter which should dynamically give the correct normalisation for both longitudinal and the transverse polarisations component up to a prefactor that mimics the missing higher order Fock components.

\begin{figure}[t]       
\includegraphics[width=0.5\textwidth,clip]{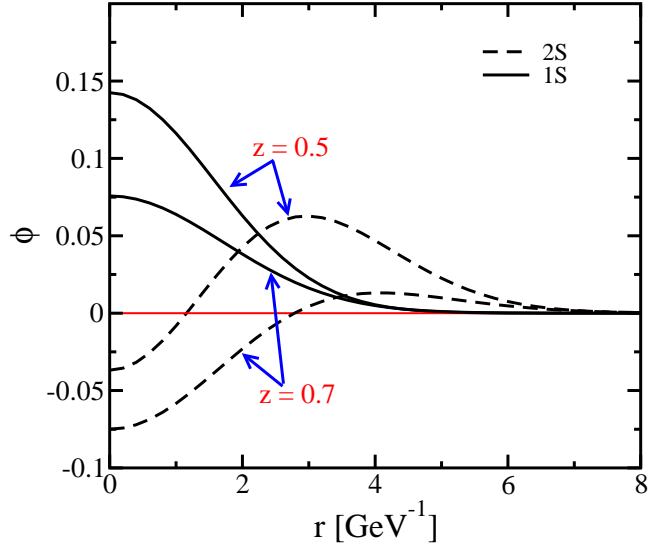}
\caption{The scalar part of the light-cone wave function of $J/\psi$  and $\psi(2s)$  with $m_c=1.4$ GeV for two different values of $z$.  }
  \label{f-phi}
\end{figure}

The parameters for $J/\psi$, $\psi(2s)$  and $\rho$ determined from the above conditions are given in table \ref{t-1}.  In this table we also compare the value of $\Gamma_{e^+e^-}$ obtained from our fit with the experimental result  $\Gamma^{exp}_{e^+e^-}$.   Note that in order to estimate the possible theoretical uncertainties associated with the condition $\mathcal{N}_{T}=\mathcal{N}_{L}$, in table \ref{t-1}, we also give a parameter set extracted by relaxing this condition. The preferred values of $\mathcal{N}_{T}$ and $\mathcal{N}_{L}$ are similar as we expect. It is also shown in table \ref{t-1} that a different value for the charm and light quark masses mainly affects the normalisation of the wave function. In \fig{f-phi}, we show the scalar part of the light-cone wave function of $J/\psi$  and $\psi(2s)$  using the parameter set corresponding to $m_c=1.4$ GeV. The position of node in $\psi(2s)$ wave function  changes with  the value of $z$. 

Note that in Ref.\,\cite{ip-sat1} it was assumed that $f_{V,T}= f_{V,L}$ while running  $\mathcal{N}_{T}$ and $\mathcal{N}_{L}$ freely in a fit. We do not impose this condition here, although the values of $f_{V,T}$ and $ f_{V,L}$ obtained  in our scheme become rather similar.  In our approach, for the case of $J/\psi$ and $\rho$, we obtained $\Gamma_{e^+e^-}=5.54$ KeV and $7.02$  KeV  while in the approach of Ref.\,\cite{b-cgc-a}  for the same quark masses we have $\Gamma_{e^+e^-}=6.79$ and $9.52$ KeV respectively, compared to the experimental value of $\Gamma^{exp}_{e^+e^-}=5.55\pm 0.14$ for $J/\psi$ and  $\Gamma^{exp}_{e^+e^-}=7.04\pm 0.06$ for $\rho$ \cite{pdg-2012}. We checked that with the new parameter sets given in table \ref{t-1}, the description of the diffractive  $J/\psi$ production at HERA and the LHC will be similar compared to the one with the old vector meson wave function parameter sets.

%%%%%%%%%%%%%%%%%%%%%%%%%%%%%%%%%
\begin{table}
\centering
\begin{tabular}{c|c|c|c|c|c||c|c|c}
\hline\hline
Meson & $m_f/GeV$ & $\mathcal{N}_L$ & $\mathcal{N}_T$ & $\mathcal{R}^2$/$\text{GeV}^2$ & $\alpha_{2s}$ & $M_V$/GeV & $\Gamma^{exp}_{e^+e^-}$/KeV &$\Gamma_{e^+e^-}$/KeV \\ \hline
$J/\psi$ & $m_c=1.4$& $0.57$ & $0.57$ & $2.45$ & $0$ & $3.097$ &$5.55\pm 0.14$ &$5.54$ \\ \hline
$J/\psi$ & $m_c=1.27$& $0.592$ & $0.596$ & $2.45$ & $0$ & $3.097$ &$5.55\pm 0.14$ &$5.46$ \\ \hline
$\psi(2s)$ &$m_c=1.4$& 0.67 &0.67 & 3.72& -0.61& 3.686& $2.37\pm 0.04$& 2.39\\ \hline
$\psi(2s)$ &$m_c=1.27$& 0.69 &0.70 & 3.72& -0.61& 3.686& $2.37\pm 0.04$& 2.35\\ \hline
$\rho$ & $m_{u,d,s}=0.01$& 0.894 & 1.004& 13.3& 0&0.775 & $7.04\pm 0.06$ & 7.06\\ \hline
$\rho$ & $m_{u,d,s}=0.14$ & 0.852& 0.908& 13.3& 0&0.775 & $7.04\pm 0.06$ & 7.02\\ \hline
\end{tabular}
\caption{Parameters of the boosted Gaussian vector meson wave functions for $J/\psi$, $\psi(2s)$ and $\rho$ obtained for two different values of quark masses. }
\label{t-1}
\end{table}
%%%%%%%%%%%%%%%%%%%%%%%%%%%%%%%%

%%%%%%%%%%%%%%%%%%%%%%%%%%%%%%%%
%%%%%%%%%%%%%%%%%%%%%%%%%%%%%%%%
%%%%%%%%%%%%%%%%%%%%%%%%%%%%%%%%
%%%%%%%%%%%%%%%%%%%%%%
\section{Impact-parameter dependent dipole models}
The common ingredient for the total (and reduced) cross-sections (i.e. for the proton structure functions in DIS) and for exclusive diffractive vector meson production \eq{am-i},  is the universal $q\bar{q}$ dipole-proton forward scattering amplitude.  Although the impact-parameter dependence of the dipole amplitude is less important for inclusive processes, it is crucial for describing  exclusive diffractive ones.  Note that the impact-parameter profile of the dipole amplitude entails intrinsically non-perturbative physics,  which is beyond the QCD weak-coupling approach to  small-$x$ physics  \cite{al,bk-c,ana}. 
Therefore, the impact-parameter dependence of the dipole amplitude, unfortunately, can only be treated phenomenologically at this time.
Supported by experimental data, it is generally assumed  a Gaussian profile for gluons where the width of the profile, as only  free parameter, is fixed via a fit to diffractive data at HERA. We use two well-known impact-parameter dependent saturation models, the so-called IP-Sat \cite{ip-sat-a,ip-sat0} and b-CGC \cite{watt-bcgc, b-cgc-a} models which both have been very successful in phenomenological applications from HERA to RHIC and the LHC.  

In the IP-Sat model \cite{ip-sat0}, the proton-dipole forward scattering amplitude is given by
%\begin{align*}
\bea
\mathcal{N}\left(x,r,b\right)  &=&1-\exp\left(-\frac{\pi^{2}r^{2}}{2N_{c}}\alpha_{s}\left(\mu^{2}\right)xg\left(x,\mu^{2}\right)T_{G}(b)\right)\,, \label{ip-sat} \\
T_{G}(b)&=& \frac{1}{2\pi B_G}\exp\left(-b^2/2B_G\right) \label{ip-b}, \
\eea
where $T_G(b)$ is the gluon impact-parameter profile and  $xg\left(x,\mu^{2}\right)$ is the gluon density, evolved with dipole transverse size $r$ up to  the scale $\mu^{2}=4/r^{2}+\mu_{0}^{2}$ with LO DGLAP gluon evolution (neglecting its coupling to quarks)  with  
initial gluon distribution at the scale $\mu_0^2$
\bea
 xg\left(x,\mu_{0}^{2}\right) &=&A_{g}\,x^{-\lambda_{g}}(1-x)^{5.6} \label{g}.\
\eea
We take the corresponding one-loop running-coupling value of $\alpha_s$ for four flavours, with $\Lambda_{\text{QCD}}=0.156$ GeV fixed by the experimentally measured value of $\alpha_s$ at the $Z^0$ mass. The contribution from bottom quarks is neglected. The IP-Sat dipole amplitude can be derived at the classical level in the CGC~\cite{mv}. Through eikonalization  it explicitly maintains unitarity while matching smoothly  the high $Q^2$ perturbative QCD limit via  DGLAP evolution. The eikonalization of the gluon distribution in the IP-Sat model represents a resummation of higher twist contributions  which become important at small $x$. The first term of the expansion of the exponential in \eq{ip-sat} corresponds to the leading-order pQCD expansion for the dipole amplitude in the colour-transparency region, as opposed to the saturation case, and it is here called the 1-Pomeron model.  

In the b-CGC dipole model \cite{watt-bcgc}, the colour dipole-proton forward scattering amplitude is given by
\bea \label{CA5}
\mathcal{N}\Lb x, r, b\Rb\,\,=\,\, \left\{\begin{array}{l}\,\,\,N_0\,\Lb \frac{r Q_s}{2}\Rb^{2\gamma_{eff}}\ \ {\rm for } \ \ r Q_s\,\leq\,2\,,\\ \\
1\,\,-\,\,\exp\Lb -\mathcal{A} \ln^2\Lb \mathcal{B} r Q_s\Rb\Rb\ \ {\rm for} \ \ \ rQ_s\,>\,2\,,\end{array}
\right.
\eea
where the effective anomalous dimension $\gamma_{eff}$ and the saturation scale $Q_s$ of the proton explicitly depend on the impact parameter and are defined as 
\bea \label{g-eff}
\gamma_{eff}&=&\gamma_s\,\,+\,\,\frac{1}{\kappa \lambda Y}\ln\Lb\frac{2}{r Q_s}\Rb, \nonumber\\
Q_s\equiv Q_{s}(x,b)&=&\Lb\frac{x_0}{x}\Rb^{\frac{\lambda}{2}}\,\exp\left\{- \frac{b^2}{4\gamma_s B_{CGC}}\right\} \text{GeV}, \
\eea
where  $Y=\ln(1/x)$ and $\kappa= \chi''(\gamma_s)/\chi'(\gamma_s)$, with $\chi$ being the LO BFKL
characteristic function. The parameters $\mathcal{A}$ and
$\mathcal{B}$ in \eq{CA5} are determined uniquely from the matching of the dipole amplitude and
its logarithmic derivatives at $r Q_s=2$.  The b-CGC model is constructed by smoothly interpolating between two analytically known limiting cases \cite{IIM}, namely the solution of the BFKL equation in the vicinity of the saturation line for small dipole sizes, and the solution of the BK equation deep inside the saturation region for large dipole sizes \cite{LT,lt3}. 

Although both the b-CGC and the IP-Sat models include saturation effects and depend on impact-parameter, 
the underlying dynamics of two models is quite different, namely saturation in the b-CGC and the IP-Sat models is probed through the increase of the gluon density (in the dilute regimes) driven by BFKL and  DGLAP evolutions, respectively. For detailed comparisons of two saturation models, see Ref.\,\cite{b-cgc-a}.

The parameters of the dipole amplitudes in the IP-Sat ($\mu_0, A_g, \lambda_g$) and  b-CGC  ($N_0$, $\gamma_s, x_0, \lambda$) models were determined via a fit to the recent combined HERA data for the reduced cross-sections \cite{Aaron:2009aa,Abramowicz:1900rp}  in the range $Q^2\in [0.75, 650]\, \text{GeV}^2$ and $x\le 0.01$.  The widths of the impact-parameter profiles, $B_{G}$ and $B_{CGC}$ in the IP-Sat and b-CGC models respectively, were iteratively fixed to give a good description of the $t$-dependence of exclusive diffractive $J/\psi$ production at HERA (at small-$t$ where data lie), while at the same time this consistently fixes the normalisation of the inclusive reduced cross-section without further adjustment  and give an excellent description of all other diffractive data (for different vector mesons and DVCS production) at small $x$ \cite{ip-sat-a,b-cgc-a}. The values of parameters of the models can be found in Refs.\, \cite{ip-sat-a,b-cgc-a}.  Note that in both the IP-Sat and b-CGC models, the fit to the recent combined HERA data at $x\le 0.01$ becomes stable for $Q^2 \ge Q^2_{min} =0.75\, \text{GeV}^2$: one observes a steady increase in $\chi^2$ with decreasing values of $Q^2_{min}$ \cite{ip-sat-a,b-cgc-a}. Therefore, our photoproduction results at $Q\approx 0$ may be considered as a test of the model beyond the kinematics where it was fitted. But the generic features of our results there, are expected not to be affected by this extrapolation.

For vector meson production, the dipole amplitude in \eq{am-i}  is evaluated at $x=x_{Bj}\left(1+M^2_V/Q^2\right)$, where $M_V$ denotes the mass of the vector meson\footnote{At $Q^2=0$, we have $x=M^2_V/(W_{\gamma p}^2-M^2_N)$ where $M_N$ denotes the nucleon mass and $W_{\gamma p}$ is the center-of-mass energy of the photon-proton system.} and $x_{Bj}$ is Bjorken-$x$. 

We stress again that in the master equations (\ref{am-i}), (\ref{vm}), (\ref{eq:beta}), the small-$x$ dynamics encoded in the dipole amplitude $\mathcal{N}\Lb x, r, b\Rb$, including its impact-parameter dependence, is the same for different vector mesons $J/\psi, \psi(2s), \phi,\rho$ and for DVCS, while the overlap wave functions between the photon and the vector mesons $\Psi_{V}^{*}\Psi$, control the typical transverse dipole size which contributes at a given kinematics.

\section{Main numerical results and predictions}
\subsection{Total cross-section of exclusive diffractive production }
\begin{figure}[t]       
\includegraphics[width=0.55\textwidth,clip]{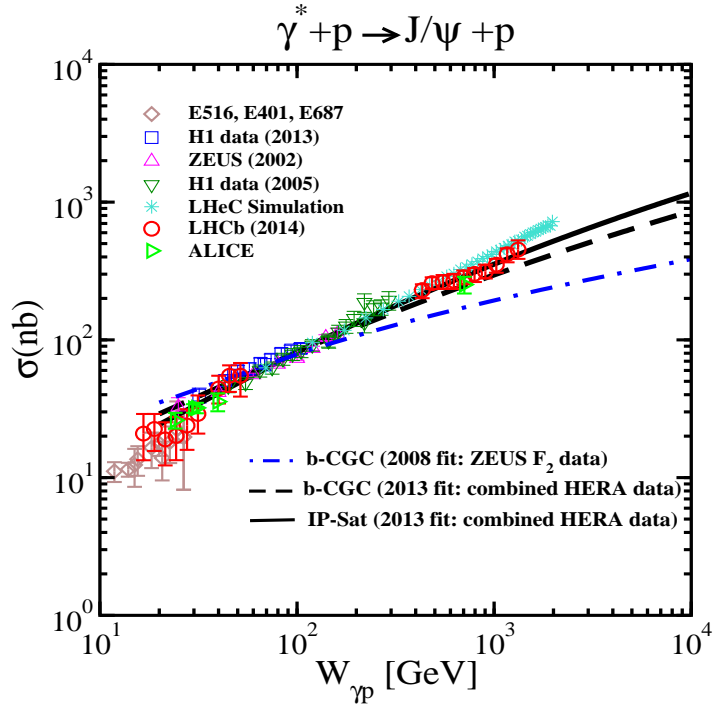}
\caption{Total $J/\psi$ cross-section as a function of $W_{\gamma p}$, compared to results from the b-CGC and IP-Sat models with parameters of the models determined  via a fit to the recent combined data from HERA \cite{ip-sat-a,b-cgc-a} and the old $F_2$ structure function \cite{watt-bcgc}  (dashed-dotted line, labeled b-CGC 2008).
  The data are from fixed target experiments \cite{fix}, the H1, ZEUS \cite{h1-2013,Chekanov:2002xi,Chekanov:2004mw,Aktas:2005xu}, LHCb \cite{lhcbn} and ALICE  (preliminary data) \cite{TheALICE:2014dwa} Collaborations. We also show the LHeC pseudo-data obtained from a simulation \cite{lhec}.}
  \label{f-vw1-ad}
\end{figure}    
\begin{figure}[t]       
\includegraphics[width=0.6\textwidth,clip]{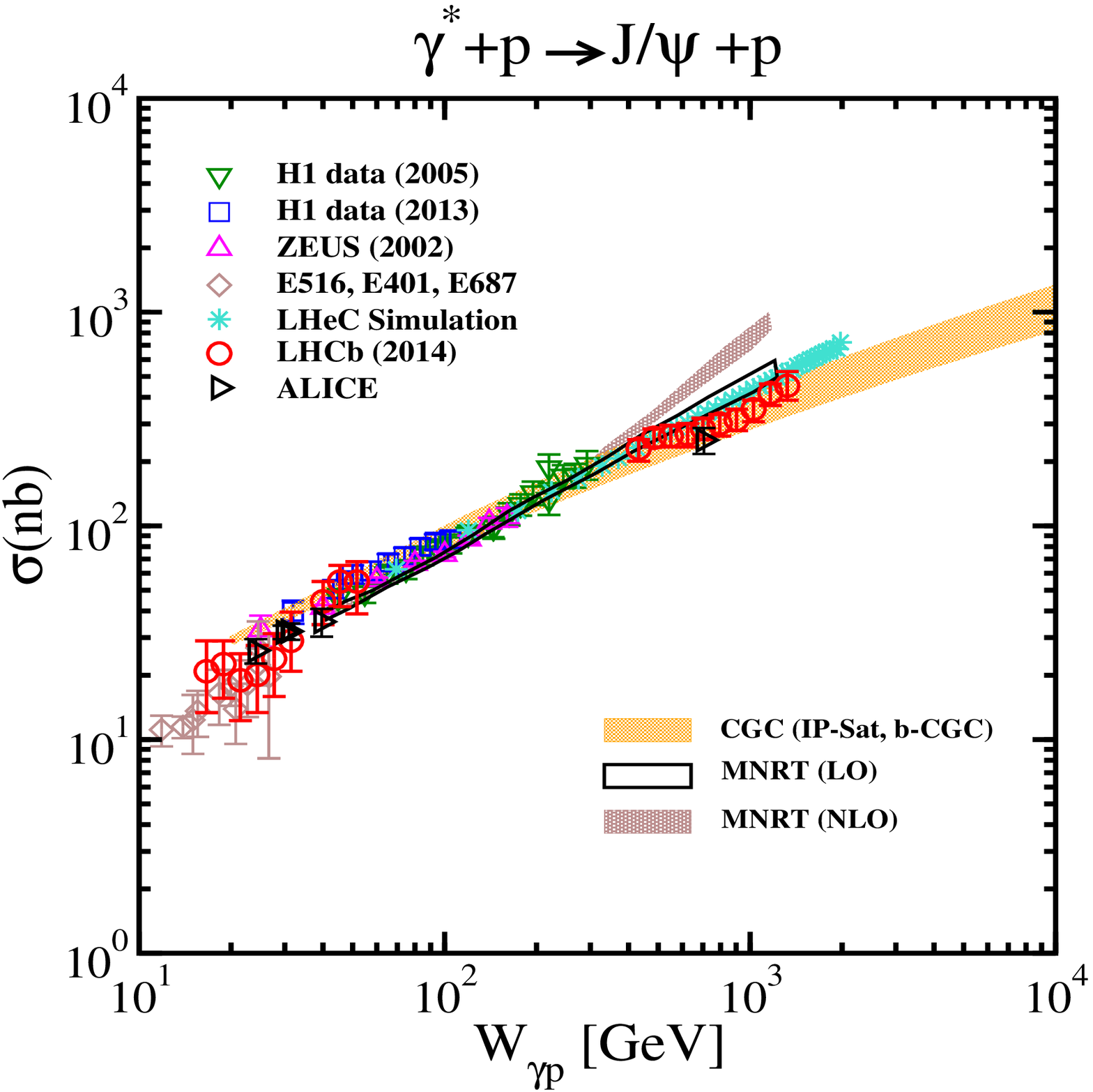}
\caption{Total $J/\psi$ cross-section as a function of $W_{\gamma p}$, compared to results from the CGC/Saturation (orange band) calculated from the b-CGC and IP-Sat models \cite{ip-sat-a,b-cgc-a}. The CGC band includes the uncertainties associated with our freedom to choose the charm mass within the range $m_c= 1.2 \div 1.4$ GeV. The results of pQCD fits at LO and NLO  \cite{pqcd-j} are taken from \cite{h1-2013}. The experimental data are the same as in \fig{f-vw1-ad}. }
  \label{f-vw1}
\end{figure}   

We first focus on the total cross-section of elastic diffractive  production of various vector mesons.   Here and thereafter, for the total cross-section we perform the integral over $|t|\in[0,1]\,\text{GeV}^2$ (unless it is explicitly given).  The advantage of the $J/\psi$  over other vector mesons is that because of its large mass, the calculation both for the cross-section and the overlap wave function are under better theoretical control and can be treated perturbatively. 
In \fig{f-vw1-ad}, we compare the results  for the total $J/\psi$ cross-section as a function of center-of-mass energy of the photon-proton system $W_{\gamma p}$, obtained using the  IP-Sat and b-CGC dipole models with a fixed charm mass $m_c=1.27$ GeV. Both models with parameters extracted via a fit  to the recent combined HERA data \cite{ip-sat-a,b-cgc-a}, give consistent results with the LHCb data \cite{lhcb}.  However, the b-CGC model with the parameters extracted via a fit to the old data (the $F_2$ structure function) \cite{watt-bcgc}, underestimates the recent LHCb data. The results obtained from the b-CGC and IP-Sat models are slightly different at very high energies due to the fact the power-law behaviour of the saturation scale in these two models is different \cite{b-cgc-a}.

\begin{figure}[t]       
\includegraphics[width=0.49\textwidth,clip]{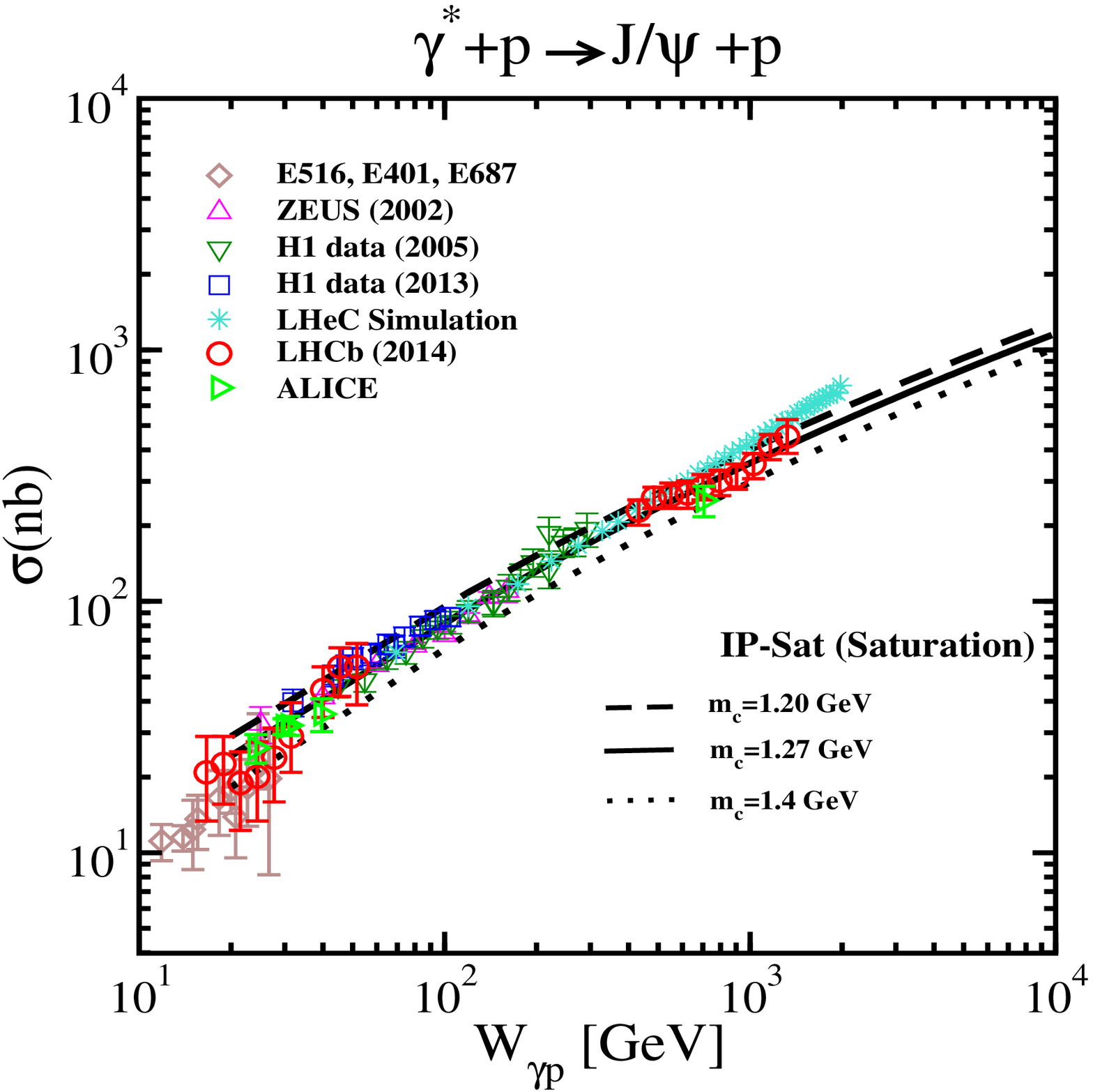}
\includegraphics[width=0.49\textwidth,clip]{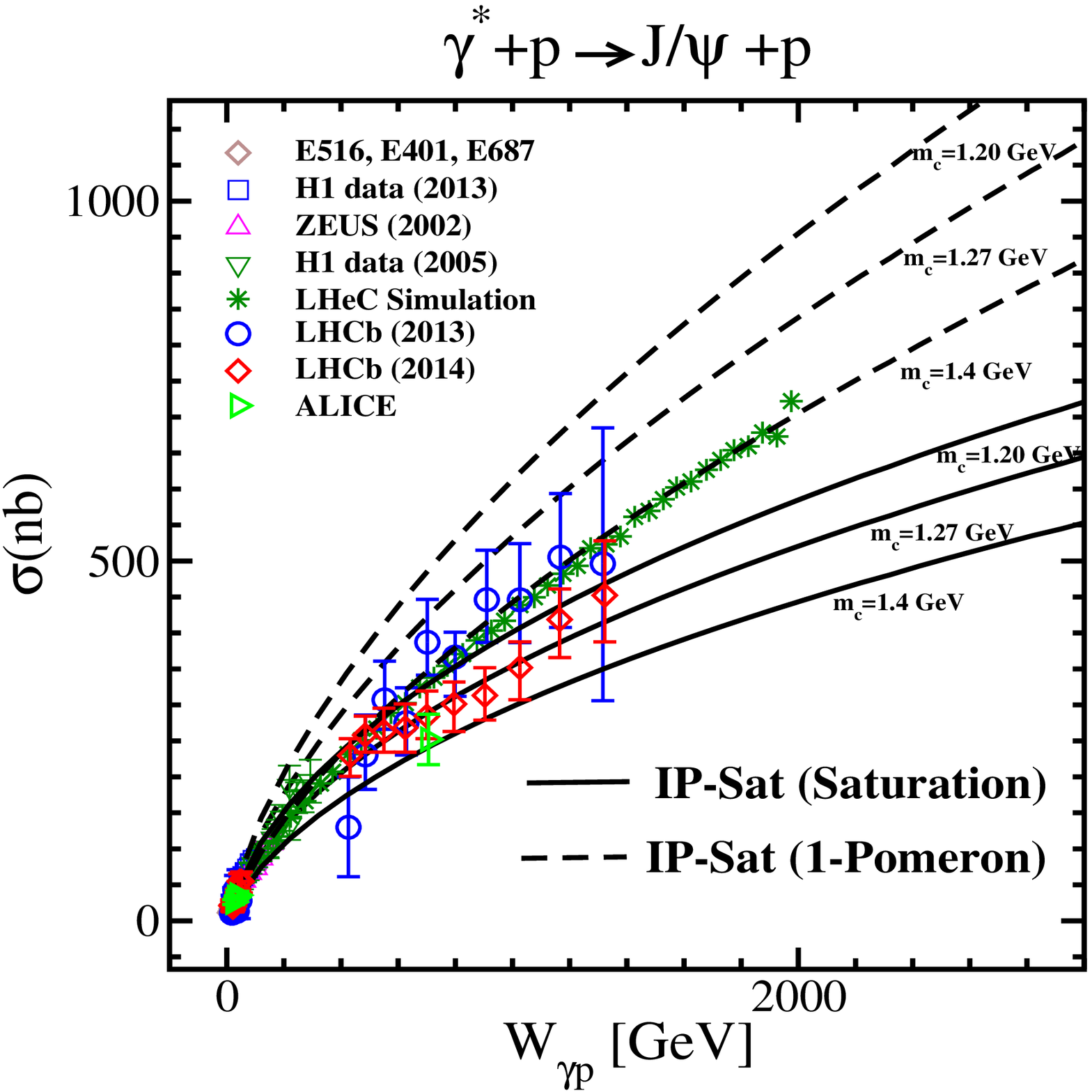}
\caption{Left: Total $J/\psi$ cross-section as a function of $W_{\gamma p}$, compared to results from the IP-Sat  model with different charm mass $m_c$. Right: Total $J/\psi$ cross-section as a function of $W_{\gamma p}$, compared to the results from the IP-Sat (saturation) 1-Pomeron models with different charm mass $m_c$. The experimental data are the same as in \fig{f-vw1-ad}.  }
  \label{f-vw2}
\end{figure}    
\begin{figure}[th]
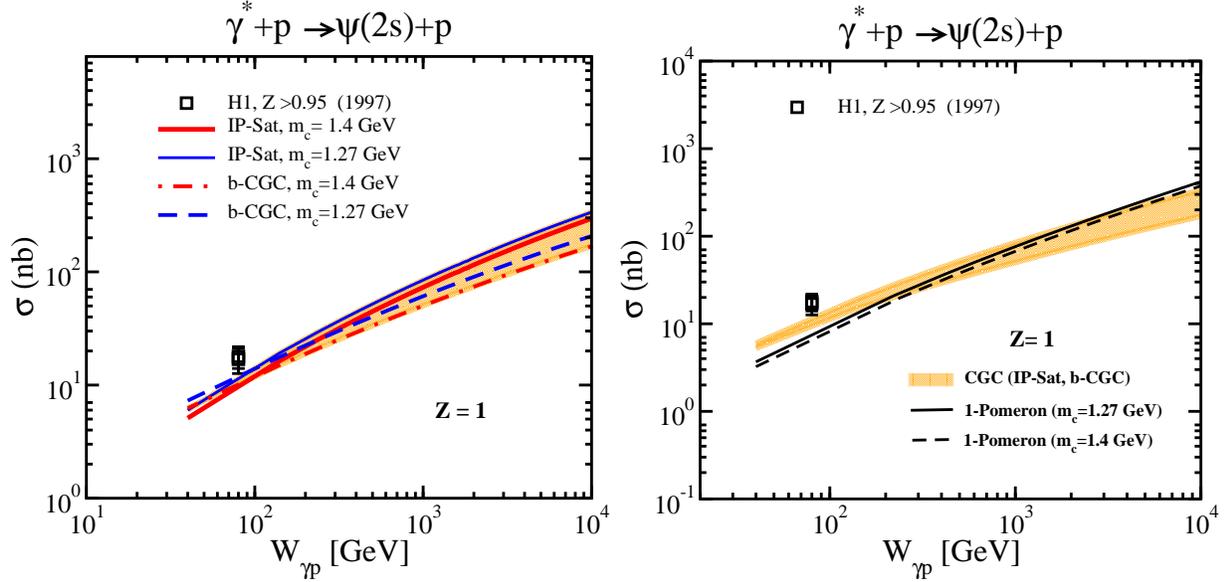
       
\includegraphics[width=0.47\textwidth,clip]{plot-psi-w-pho2.eps}
\includegraphics[width=0.47\textwidth,clip]{plot-psi-w-pho3.eps}
\caption{Left: Total $\psi(2s)$ diffractive photoproduction cross-section as a function of $W_{\gamma p}$, compared to results from the IP-Sat and  b-CGC  models with different charm mass $m_c$. Right: Similar to the left panel, the results of the CGC/saturation (orange band) and 1-Pomeron models are compared. The experimental data are from the H1 collaboration \cite{h1-psi} for quasi-elastic ($Z>0.95$) photoproduction of $\psi(2s)$ while all theory curves are for elastic diffractive production with elasticity $Z=1$. }
  \label{f-psi1}
\end{figure}

In \fig{f-vw1}, we compare the results obtained from saturation models and from a pQCD approach at LO and NLO \cite{pqcd-j} with all available data from fixed target experiments to the recent ones from the H1, ZEUS, LHCb and ALICE collaborations\footnote{For the purpose of illustrating the precision that could be achieved in future experiments, both in \fig{f-vw1-ad} and in \fig{f-vw1} we also show the LHeC pseudo-data obtained from a simulation \cite{lhec} based on a power-law extrapolation of HERA data.}  \cite{lhcbn,TheALICE:2014dwa,h1-2013,fix,Chekanov:2002xi,Chekanov:2004mw,Aktas:2005xu}.  The band labeled  "CGC" includes the saturation results obtained from the IP-Sat and b-CGC models with the parameters of models constrained by the recent combined HERA data. Note that the LHCb data points in \fig{f-vw1} were not used for fixing the model parameters, and therefore our CGC results in \fig{f-vw1} at high energy can be considered as predictions. Also note that diffractive $J/\psi$ production is sensitive to the charm quark mass at low $Q^2$. This is because the scale in the integrand of the cross-section is set by the charm quark mass for low virtualities $Q^2<m_c^2$.  The CGC band in \fig{f-vw1} also includes the uncertainties associated with choosing the charm mass within the range $m_c=1.2 \div 1.4$ GeV extracted from a global analysis of existing data at small-x $x<0.01$ \cite{ip-sat-a,b-cgc-a}.  In \fig{f-vw1}, we compare with the LHCb updated data released in 2014 \cite{lhcbn} which  are significantly more precise compared to earlier measurements \cite{lhcb} (see also \fig{f-vw2} right panel). It is seen that the ALICE \cite{TheALICE:2014dwa} and LHCb \cite{lhcbn} data are in good agreement with the CGC predictions while there seem to be some tensions between the experimental data and the pQCD results (labeled MNRT LO and NLO) at high $W_{\gamma p}$. It was recently shown that including the LHCb data in the pQCD fit, allows a better constraint on the low-$x$ gluon distribution \cite{n-pqcd}.

\begin{figure}[t]
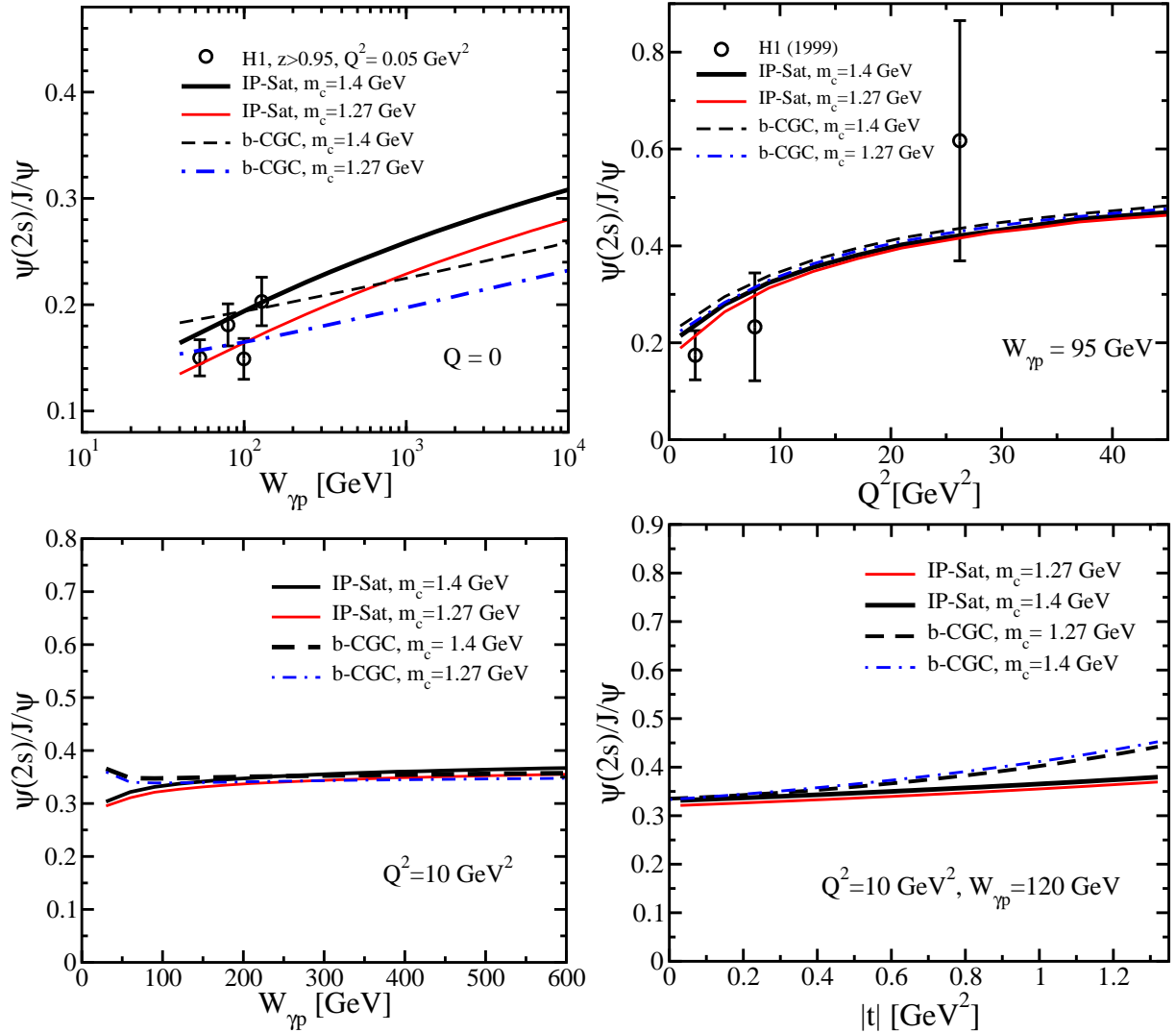
       
\includegraphics[width=0.47\textwidth,clip]{plot-psi-w-photo.eps}
\includegraphics[width=0.47\textwidth,clip]{plot-psi-q2-22.eps}
\includegraphics[width=0.47\textwidth,clip]{plot-psi-w-fixedQ2.eps}
\includegraphics[width=0.47\textwidth,clip]{plot-psi-t.eps}
\caption{ Top Left: The ratio of the cross-section for $\psi(2s)$ and $J/\psi$  ($R=\psi(2s)/J/\psi$) for diffractive photoproduction as a function of $W_{\gamma p}$. Top right: The ratio $R$ for diffractive production as a function of $Q^2$ at a fixed $W_{\gamma p}=95$ GeV. Bottom left: The ratio $R$ for diffractive production as a function of $W_{\gamma p}$ at a fixed $Q^2=10$ GeV.  Bottom right: The ratio $R$ for diffractive production as a function of $|t|$ at a fixed $Q^2=10$ GeV and  $W_{\gamma p}=120$ GeV. In all panels, the theoretical curves are the results from the IP-Sat and b-CGC models with different parameter sets corresponding to different charm masses.  The experimental data are from the H1 Collaboration \cite{h1-ratioi}.}
  \label{f-ratio}
\end{figure}  

In \fig{f-vw2}, we show the charm-mass dependence of the total $J/\psi$ cross-section as a function of $W_{\gamma p}$. 
Within the saturation models, a lower charm mass about $m_c\approx 1.27 $ GeV is preferred. However, in the non-saturation version of the IP-Sat model (1-Pomeron), a larger charm mass about $m_c\approx 1.4 $ GeV provides a better agreement with experimental data (see the right panel of that figure). In \fig{f-vw2} right panel, we also show ALICE preliminary data \cite{TheALICE:2014dwa}, the LHCb updated data (labeled LHCb 2014) \cite{lhcbn} and earlier LHCb data \cite{lhcb} (labeled LHCb 2013). It is seen that the combined ALICE and LHCb updated 2014 data are more in favour of the saturation than of the 1-Pomeron model results at high $W_{\gamma p}$.   
Nevertheless, in order to clearly discriminate among models one should first more accurately determine the charm mass. This can be done by precise measurements of the charm structure function $F^{c}_2$ or a reduced cross-section for charm production in a wider range of kinematics, including at small virtualities, than those currently available at HERA (restricted to $Q^2\ge 2.5\,\text{GeV}^2$ and $x \ge 3\times 10^{-5}$ \cite{Abramowicz:1900rp}). Such measurements can in principle be done in the projected LHeC \cite{lhec}.

\begin{figure}[t]
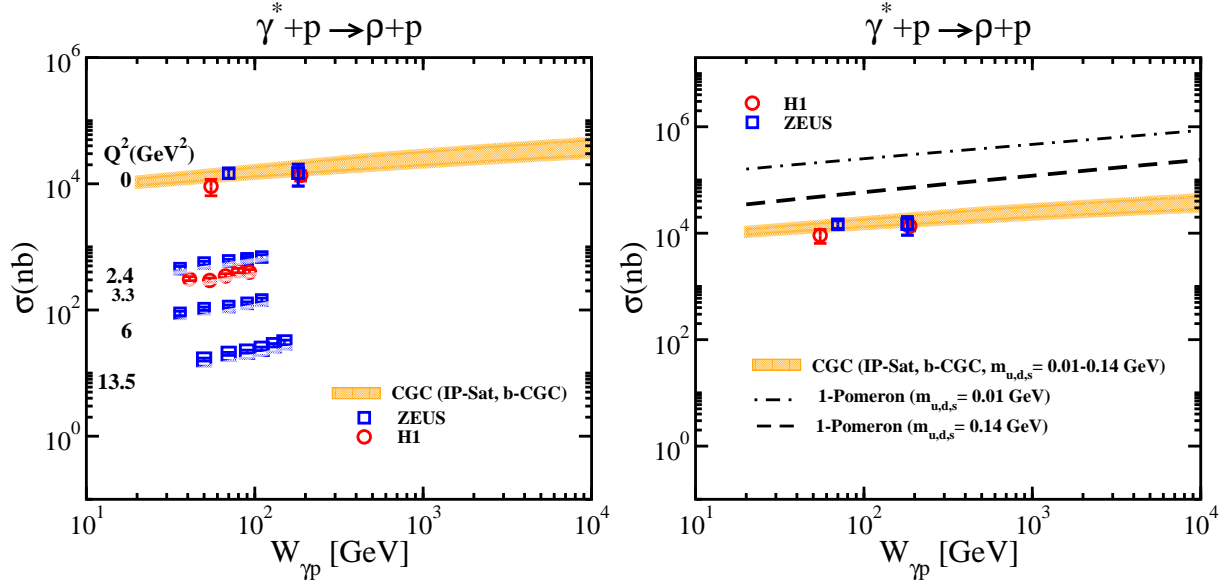
       
\includegraphics[width=0.47\textwidth,clip]{plot-rho-w-photo.eps}
\includegraphics[width=0.47\textwidth,clip]{plot-rho-w-mass.eps}
\caption{Left: Total diffractive $\rho$ cross-section as a function of $W_{\gamma p}$ at different virtualities $Q^2=0, 2.4,3.3,6, 13.5\,\text{GeV}^2$. 
The CGC band  (orange band) includes the results obtained from the b-CGC and IP-Sat models and also the uncertainties associated with our freedom to choose the light quark mass within the range $m_{u,d,s}= 0.01 \div 0.14$ GeV. Right:  Total diffractive $\rho$ cross-section compared to results from the CGC/saturation and 1-Pomeron model with two different masses $m_{u,d,s}=0.01$ and $0.14$ GeV. 
The experimental data are from \cite{h1-rho,zeus-rho}.}
  \label{f-rho1}
\end{figure}

In \fig{f-psi1}, we show the total cross-section of elastic diffractive photoproduction of $\psi(2s)$ as a function of $W_{\gamma p}$ obtained from the IP-Sat and b-CGC saturation models with different charm masses corresponding to different parameter sets of the dipole amplitude.  Note that the experimental data  \cite{h1-psi} are for quasi-elastic ($Z>0.95$) photoproduction of $\psi(2s)$ while all theory curves are for elastic diffractive production with elasticity $Z=1$. The elasticity is defined as $Z=E_{\psi(2s)}/E_{\gamma}\approx (W^2-M_Y^2)/(W^2-m_p^2)$ where $M_Y$ is the effective mass of the hadrons produced in the dissociation of the proton.
In the right panel, we compare the results obtained from the 1-Pomeron and the saturation models. It is seen that  within theoretical uncertainties associated with charm mass, the 1-Pomeron and the saturation models give rather similar results in the range of energy shown in \fig{f-psi1}. This is mainly due to the fact that the  $\psi(2s)$ is heavier than $J/\psi$, therefore effective dipole sizes  $r\sim 1/\epsilon$ which contribute to the total cross-section are smaller for $\psi(2s)$ than for $J/\psi$. Note that although the scalar part of the $\psi(2s)$ wave function extends to large dipole sizes (see \fig{f-phi}), due to the existence of the node, there is large cancellation between dipole sizes above and below the node position. As a result, the total cross-section of $\psi(2s)$ is suppressed compared to $J/\psi$ production, see Figs.\,\ref{f-psi1},\ref{f-ratio}. 

In \fig{f-ratio}, we show the ratio of the cross-section for $\psi(2s)$ and $J/\psi$ for diffractive production $R=\psi(2s)/J/\psi$ as functions of $W_{\gamma p}$ at $Q=0$ (top left panel),  $Q^2$ at a fixed $W_{\gamma p}=95$ GeV (top right panel), $W_{\gamma p}$ at a fixed $Q^2=10$ GeV$^2$ (bottom left panel) and $|t|$ at a fixed $Q^2=10$ GeV$^2$ and  $W_{\gamma p}=120$ GeV (bottom left panel).  It is seen that at a fixed high virtualities, the ratio $R$ has little dependence to $|t|$ and $W_{\gamma p}$ (bottom panel), while the ratio $R$ increases with virtualities at a fixed  $W_{\gamma p}$ (top right panel). 
It is also seen in  \fig{f-ratio} (top left panel) that the photoproduction ratio $R (Q=0)$ increases with $W_{\gamma p}$ and becomes sensitive to different saturation models.  Therefore, precise measurements of the ratio of  diffractive photoproduction of $\psi(2s)$ and $J/\psi$ at HERA and the LHC can provide valuable extra constraint on the saturation models.

In \fig{f-rho1}, we show total diffractive $\rho$ meson cross-section as a function of $W_{\gamma p}$ at different virtualities $Q^2=0, 2.4,3.3,6, 13.5\,\text{GeV}^2$, compared to results obtained from the b-CGC and the IP-Sat models. In the case of photoproduction, similar to experimental measurement, we perform the integral over $t\in [0,0.5]\,\text{GeV}^2$. The orange band labeled CGC includes results from both the IP-Sat and  b-CGC models with uncertainties associated to our freedom to choose different light-quark masses within a range $m_{u,d,s}= 0.01 \div 0.14$ GeV.  We also compare the CGC/saturation results with those obtained from the 1-Pomeron model with two different light quark masses  $m_{u,d,s}= 0.01$ and $0.14$ GeV.  It is seen that 1-Pomeron results are significantly different from the saturation models, and HERA data can already rule out the 1-Pomeron model with light quark masses. Notice that increasing the light quark masses to  $m_{u,d,s}\approx 0.35 \div 0.4 $ GeV  (not shown in \fig{f-rho1}), significantly reduces the cross-section in the 1-Pomeron model and brings it closer to the saturation results with $m_{u,d,s}= 0.01 \div 0.14$ GeV. However, a dipole model with such a large light-quark masses does not provide a good description of the structure functions at very low virtualities \cite{ip-sat-a,b-cgc-a}.  
This may indicate the existence of large non-linear effects for the diffractive photoproduction of the $\rho$ meson.  Note that, as we already pointed out, the effective dipole size which contributes to the cross-section is proportional to the inverse of the meson mass at $Q=0$. Therefore the total diffractive cross-section of lighter vector meson such as the $\rho$ meson should be a better probe of saturation physics (see also below). 

\subsection{$t$-distribution of the diffractive production off protons and the origin of dips}

\begin{figure}[t]       
\includegraphics[width=0.49\textwidth,clip]{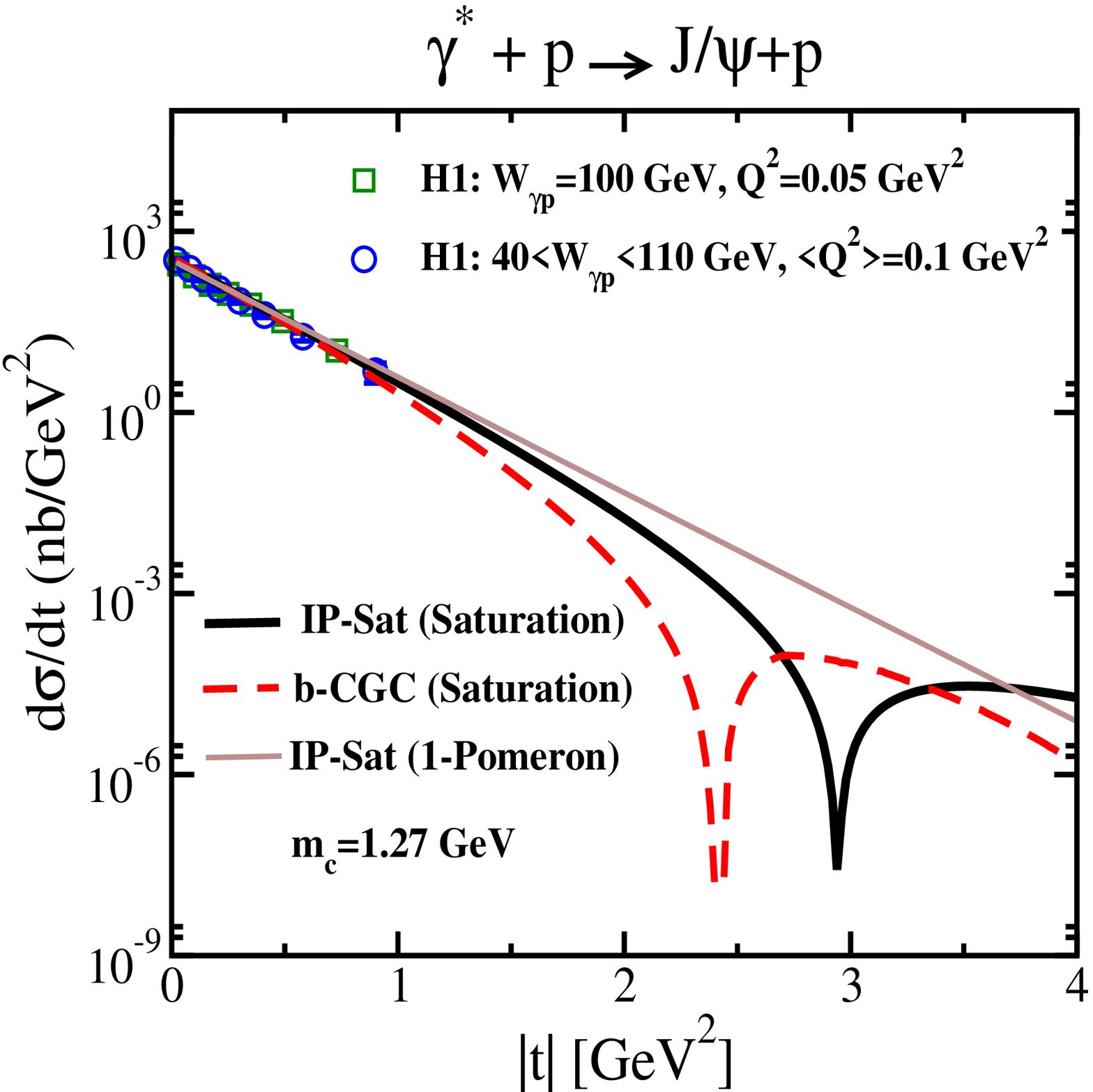}
\includegraphics[width=0.49\textwidth,clip]{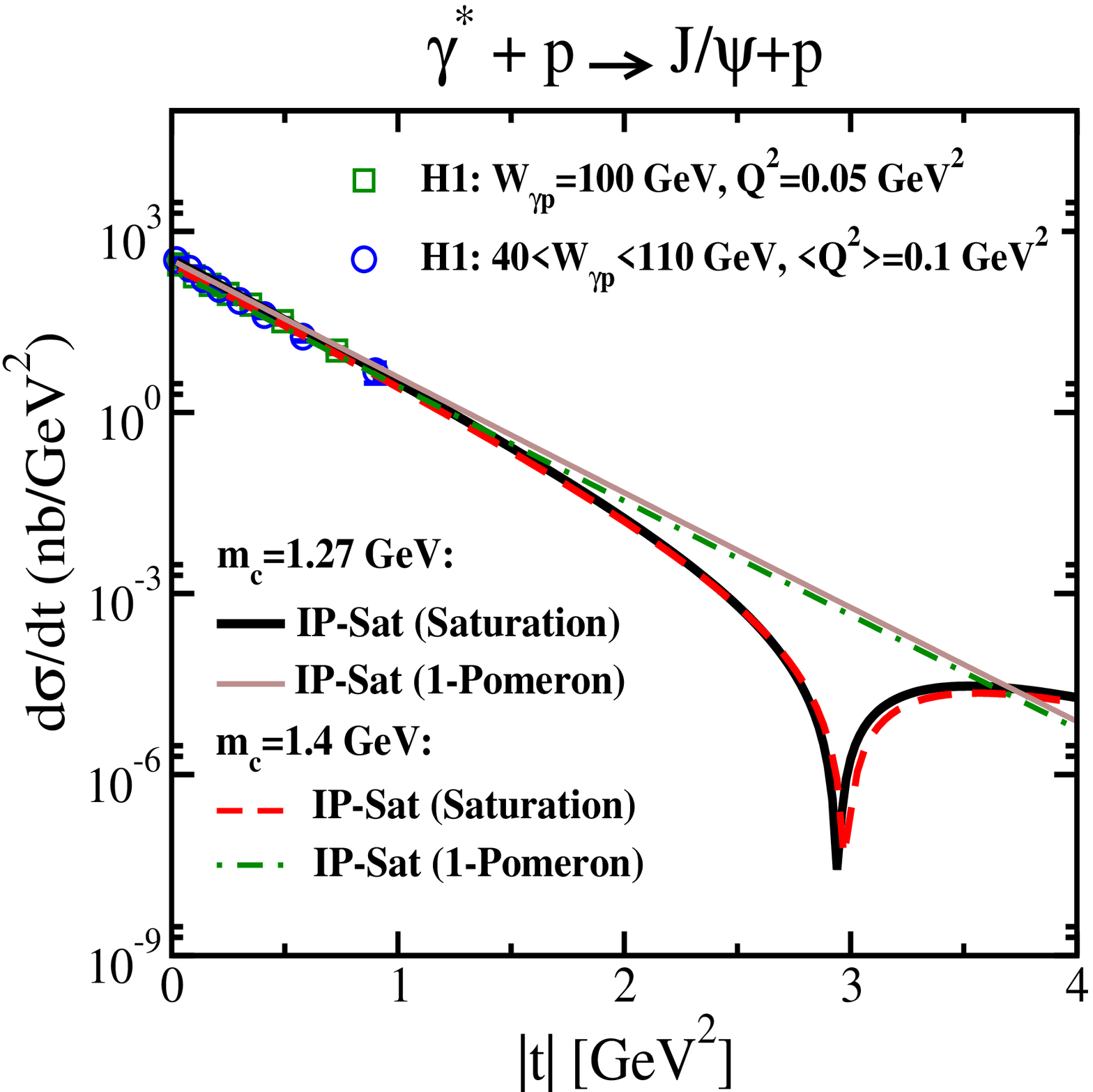}
\caption{Left: Differential vector meson cross-sections for $J/\psi$, as a function of $|t|$ within the IP-Sat, b-CGC and 1-Pomeron models with a fixed $m_c=1.27$ GeV at HERA.  Right: Results obtained from the IP-Sat and 1-Pomeron models are compared for two values of the charm mass $m_c=1.27, 1.4$ GeV. The experimental data are from the H1 Collaboration \cite{h1-2013,Aktas:2005xu}.  }
  \label{f-vt0}
\end{figure}

In \fig{f-vt0}, left panel, we compare the saturation and non-saturation models results for the $t$-distribution of  the exclusive  photoproduction of $J/\psi$ at $Q\approx 0$ with available data from HERA. It can be observed that at low $|t|$ where currently experimental data are available, one cannot discriminate between the saturation and non-saturation (1-Pomeron) models and all three models:  IP-Sat, b-CGC and  1-Pomeron, provide a good description. However, at large $|t|$ the models give drastically different results, namely both the IP-Sat and b-CGC saturation models produce a dip while the 1-Pomeron model does not.  In \fig{f-vt0}, right panel, we show the charm-mass dependence of the $t$-distribution of exclusive $J/\psi$ photoproduction. The appearance and position of the dip are only slightly affected by the choice of charm mass.  Therefore, in this respect, theoretical uncertainties due to the charm mass are less important for the $t$-distribution than for the total cross-section.

\begin{figure}[h]       
\includegraphics[width=0.49\textwidth,clip]{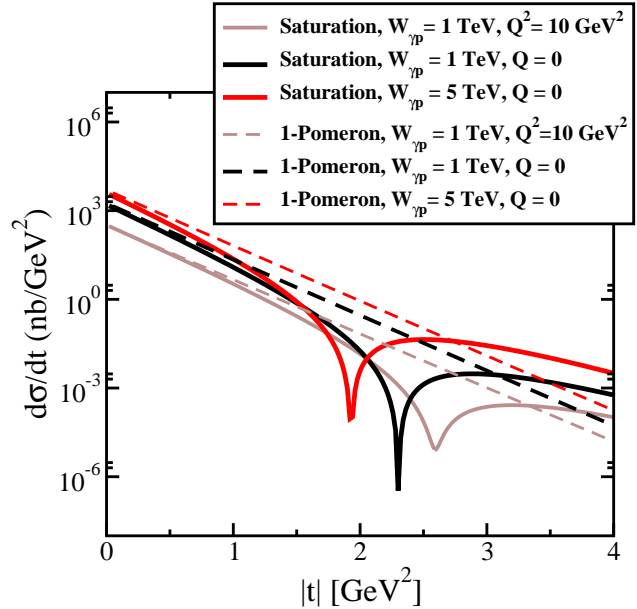}
\caption{Differential $J/\psi$ cross-section, as a function of $|t|$ within the IP-Sat (saturation) and IP-Sat (1-Pomeron) models with a fixed $m_c=1.27 $ GeV at LHC/LHeC energies $W_{\gamma p}=1, 5$ TeV and $Q^2=0, 10\,\text{GeV}^2$.  }
  \label{f-vt1}
\end{figure}    

 In \fig{f-vt1}, we show our predictions for the $t$-distribution of  exclusive $J/\psi$ photoproduction  at  LHC/LHeC energies $W_{\gamma p}=1, 5$ TeV at two virtualities $Q^2=0, 10\,\text{GeV}^2$ obtained from the IP-Sat (saturation) and the 1-Pomeron models. In the saturation model, the dip shifts to smaller values of $|t|$ for smaller $Q$ and for higher $W_{\gamma p}$. Note that saturation effects are expected to become more important at low virtualities and high energies.

\begin{figure}[t]
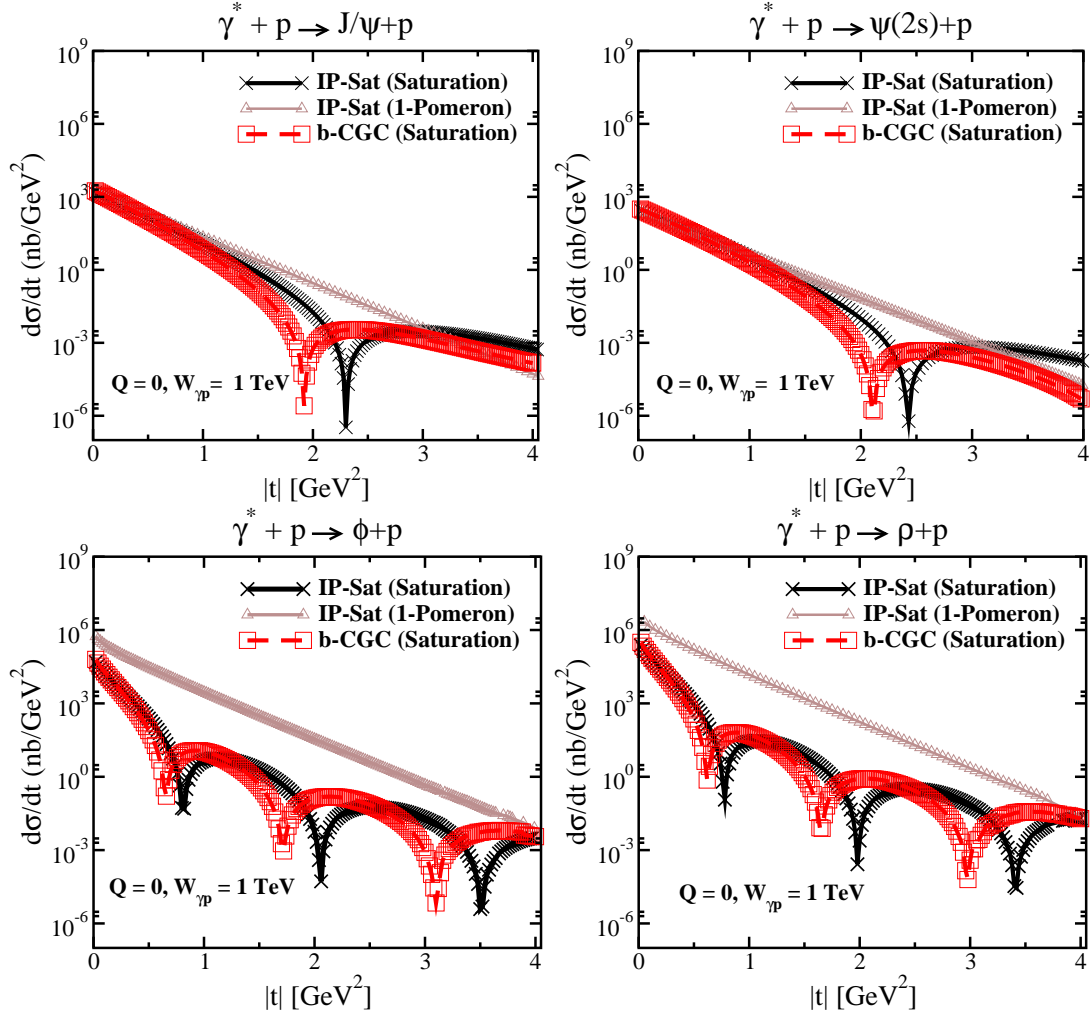
       
\includegraphics[width=0.42\textwidth,clip]{plot-jpsi-t-lhcb1.eps}
\includegraphics[width=0.42\textwidth,clip]{plot-psi-t-lhc.eps}
\includegraphics[width=0.42\textwidth,clip]{plot-phi-t-lhcb.eps}
\includegraphics[width=0.42\textwidth,clip]{plot-rho-t-lhcb.eps}
\caption{ Differential diffractive vector meson photoproduction cross-sections for $J/\psi, \psi(2s), \phi, \rho$, as a function of $|t|$ within the IP-Sat (saturation), b-CGC and 1-Pomeron models at a fixed $W_{\gamma p}=1$ TeV and $Q=0$. The thickness of points includes the uncertainties associated with our freedom to choose different values for the charm quark mass within the range $m_c \approx 1.2\div 1.4$ GeV (corresponding to different dipole parameter sets) and $m_{u,d,s}\approx 0.01$ GeV. }
  \label{f-vt2}
\end{figure}

In \fig{f-vt2}, we compare the results obtained from the IP-Sat and b-CGC models with those from the 1-Pomeron model, for the $t$-distribution of the elastic photoproduction of vector mesons\footnote{In the case of $\phi$ meson, we use boosted Gaussian wavefunction with parameters given in Ref.\cite{ip-sat1}. For other vector mesons, we use parameters for the wavefunction given in table \ref{t-1}.} $J/\psi$, $\psi(2s)$, $\phi$ and $\rho$ off the proton at an energy accessible at the LHC/LHeC, $W_{\gamma p}=1$ TeV, for $Q=0$.  Drastic different patterns for the diffractive $t$-distribution also emerge between saturation and non-saturation models for lighter vector meson production such as $\rho$ and $\phi$, with the appearance of multiple dips. Note that the prospects at the LHeC \cite{lhec} indicate that access to values of $|t|$ around 2 GeV$^2$, required to observe the dips for $J/\psi$, is challenging. On the other hand, the accuracy that can be expected at lower $|t|$ should allow to observe the bending of the distributions. And lower values of $|t|$ for lighter vector mesons should be clearly accessible, probably even at the EIC \cite{eic} but for smaller $W_{\gamma p}$.

\begin{figure}[t]       
\includegraphics[width=0.47\textwidth,clip]{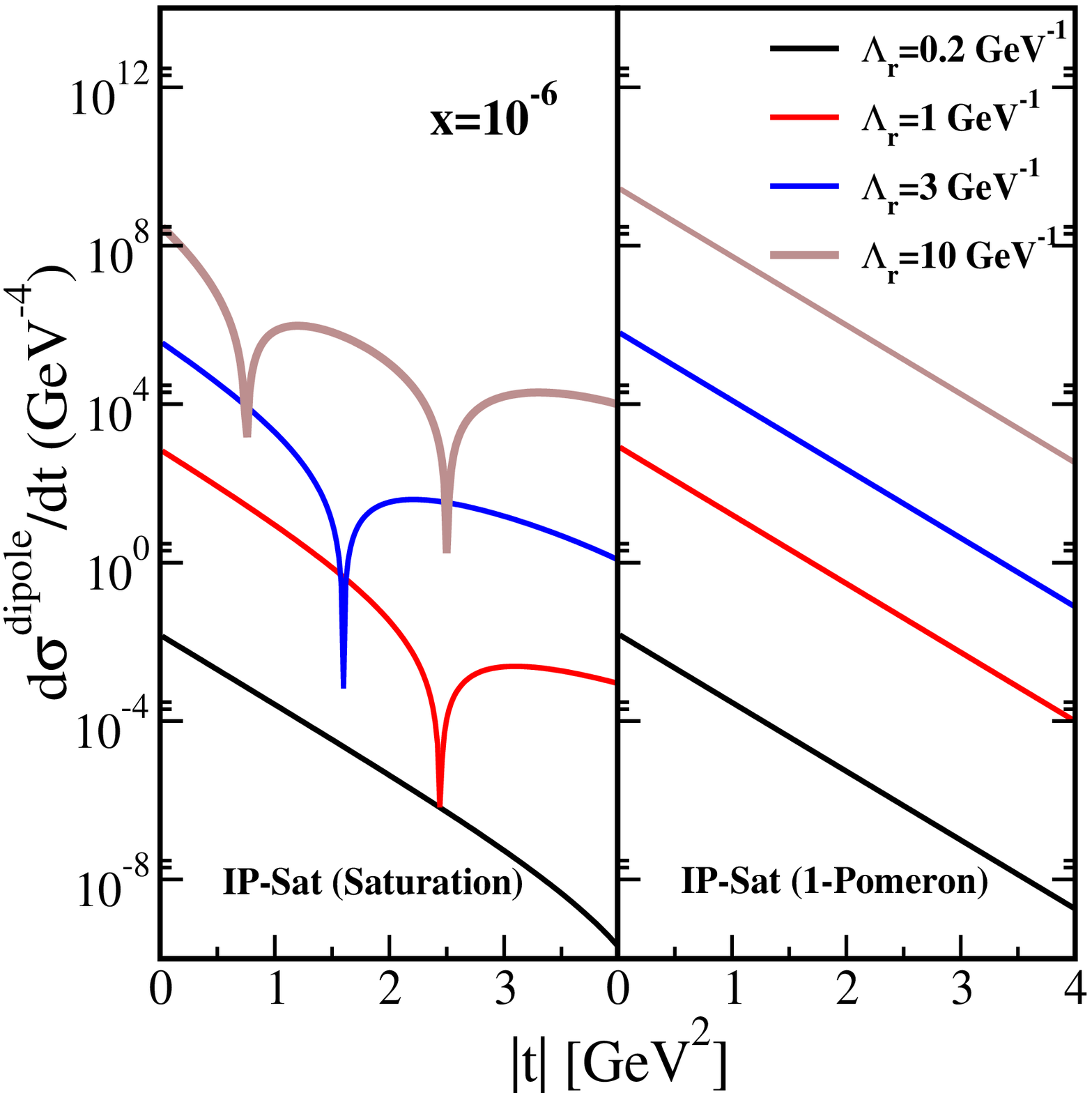}
\includegraphics[width=0.47\textwidth,clip]{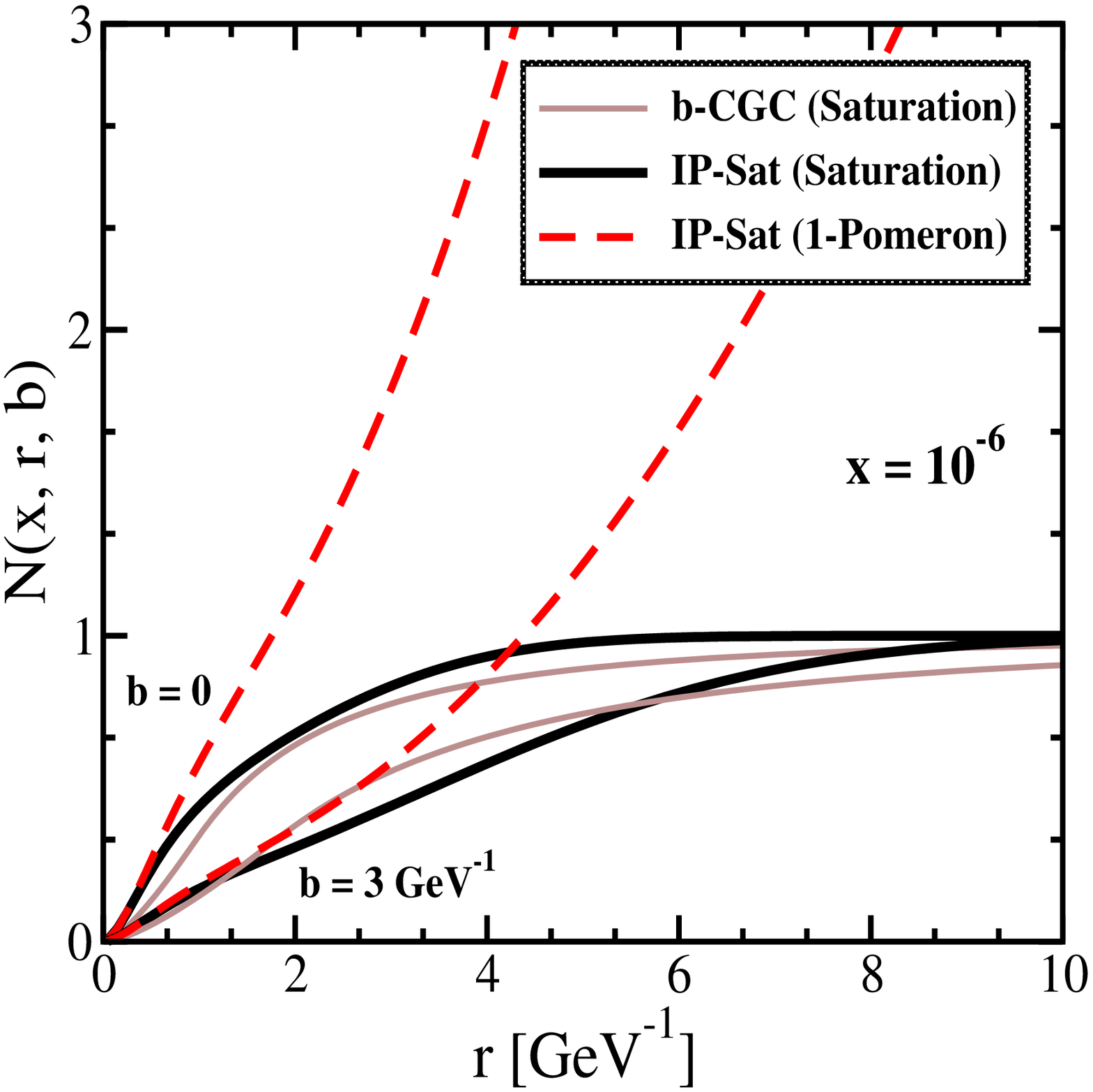}
\caption{ Left: The $t$-distribution of dipole amplitude defined in \eq{d-f} with different values of cutoff on dipole transverse-size $\Lambda_r$ in the IP-Sat (saturation) and IP-Sat (1-Pomeron) models at a fixed $x=10^{-6}$. Right: The dipole amplitude in different models as a function of dipole size $r$ at $x=10^{-6}$ for two values of impact parameter $b=0, 3\,\text{GeV}^{-1}$.  }
  \label{f-sd}
\end{figure}    

The emergence of a single or multiple dips in the $t$-distribution of the vector mesons in the saturation models is directly related to the saturation (unitarity) features of the dipole scattering amplitude $\mathcal{N}$ at large dipole sizes.  In order to see more clearly this effect, let us define a $t$-distribution of the dipole amplitude in the following way:
\begin{equation} \label{d-f}
\frac{d\sigma^{\text{dipole}}}{dt}=2\pi\Big{|}\int_{0}^{\Lambda_r} rdr\, \int \dif^2\vec{b}\, \mathrm{e}^{-\mathrm{i}\vec{b}\cdot\vec{\Delta}} \mathcal{N}\left(x,r,b\right)\Big{|}^2,
\end{equation}
where $\Lambda_r$ is an upper bound on the dipole size. The above expression is in fact very similar to Eqs.\,(\ref{am-i}), (\ref{vm}), see also Ref.\,\cite{ip-sat0}. Note that in \eq{am-i}, the overlap of photon and vector meson wave functions gives the probability for finding a colour dipole of transverse size $r$ in the vector meson wave function and it naturally gives rise to an implicit dynamical cutoff $\Lambda_r$ which varies with kinematics and the mass of the vector meson. 
The cutoff $\Lambda_r$ is larger at lower virtualities and for lighter vector mesons. On the other hand, quantum evolution leads to unitarity constrains on the amplitude at lower dipole sizes with decreasing values of $x$ or increasing energies.
Thus, by varying the cutoff $\Lambda_r$, one probes different regimes of the dipole from colour transparency to the saturation regime.

In the 1-Pomeron model, since the impact-parameter profile of the dipole amplitude is a Gaussian for all values of $r$, its Fourier transform becomes exponential for all values of $t$ irrespective of the value of the cut-off. For low $\Lambda_r$, the integrand in \eq{d-f}  is in the colour transparency regime (or the 1-Pomeron limit of the IP-Sat model), and the $b$-dependence of the amplitude is Gaussian and consequently its Fourier transform is exponential for all values of $t$. 
However,  in a case with a large cutoff $\Lambda_r$,  the typical dipole size which contributes to the integral is within the unitarity or black-disc limit, see e.g. \cite{Frankfurt:2011cs}, with $\mathcal{N}\to 1$ (see \fig{f-sd} right panel). Then, the Fourier transform of the dipole amplitude
leads to a dip or multi-dips, as seen in \fig{f-sd} (left panel).  
The saturation effect becomes more important at smaller Bjorken-$x$ or larger $W_{\gamma p}$, and lower virtualities $Q$ where the the contribution of large dipole sizes becomes more important, leading to a large effective $\Lambda_r$ and consequently  to the dip-type structure. 

For lighter vector mesons, the overlap extends to larger dipole sizes resulting in a  dip structure as seen in \fig{f-sd}. The full calculation computed from \eq{vm} and shown in \fig{f-vt2}, indeed supports the fact that lighter vector mesons (which naturally have a larger  $\Lambda_r$) develop multiple dips within the same kinematic region in which the heavier vector meson has a single dip (with a correspondingly smaller $\Lambda_r$), consistent with the expectation in the saturation picture shown in \fig{f-sd} (left panel).  The exact position of dips and whether the t-distribution has  multiple or a single minimum depend on the value of dynamical cutoff $\Lambda_r$ (via the kinematics and the mass of vector mesons) and the impact-parameter profile of the saturation scale. In the case of $\psi(2s)$ vector meson, although the scalar part of the $\psi(2s)$ wave function extends to large dipole sizes, due to the node effect, there is large cancellation between dipole sizes above and below the node position. As a result, the total cross-section of $\psi(2s)$ is suppressed compared to $J/\psi$ production as seen in \fig{f-ratio} and the dip in the $t$-distribution moves slightly to higher $|t|$ compared to diffractive $J/\psi$ production. We recall that  $\psi(2s)$ is slightly heavier than $J/\psi$ and consequently the dip (for a heavier vector meson) moves toward higher $|t|$ compared to $J/\psi$ production.

\begin{figure}[t]       
\includegraphics[width=0.47\textwidth,clip]{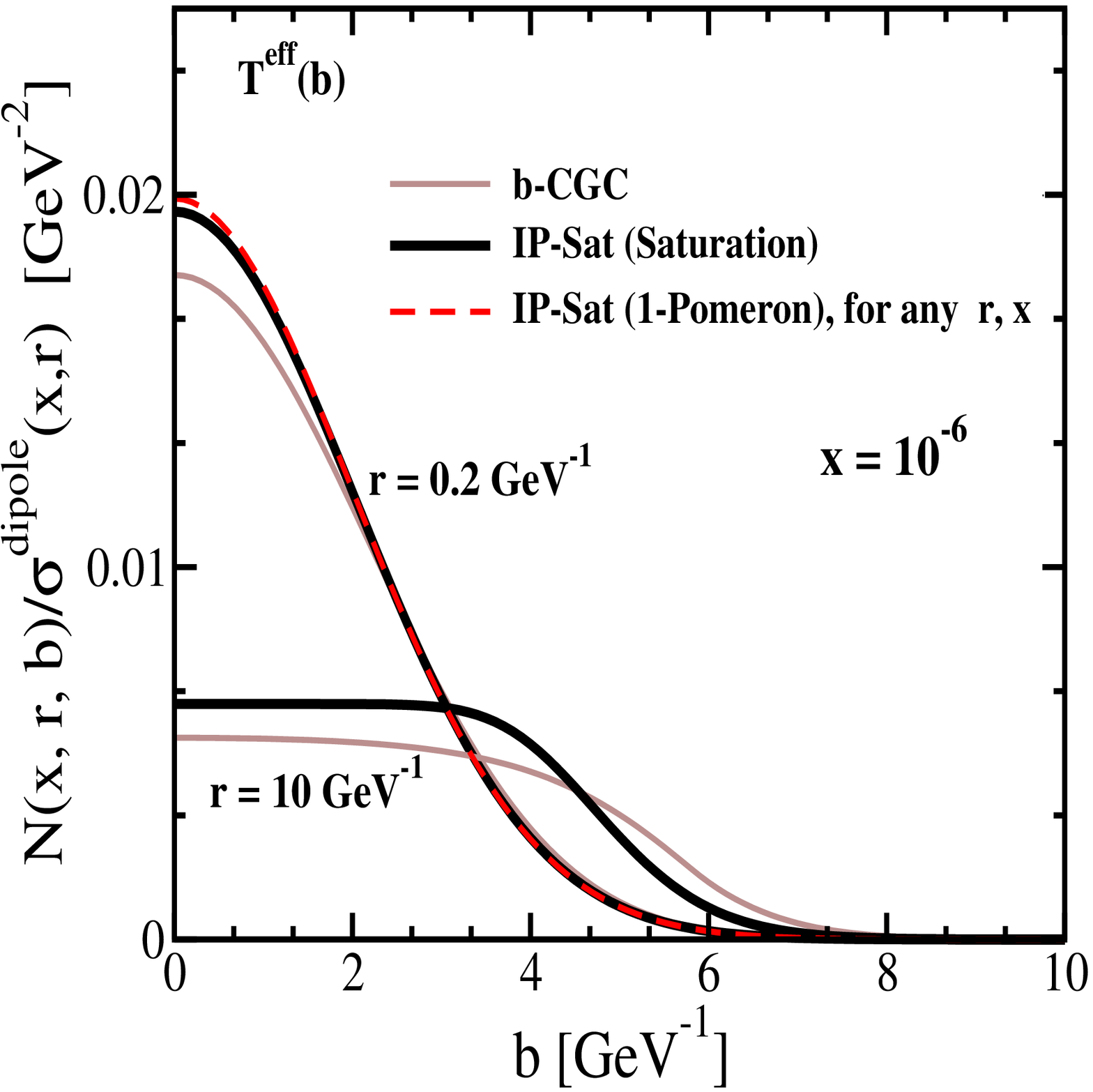}
\includegraphics[width=0.47\textwidth,clip]{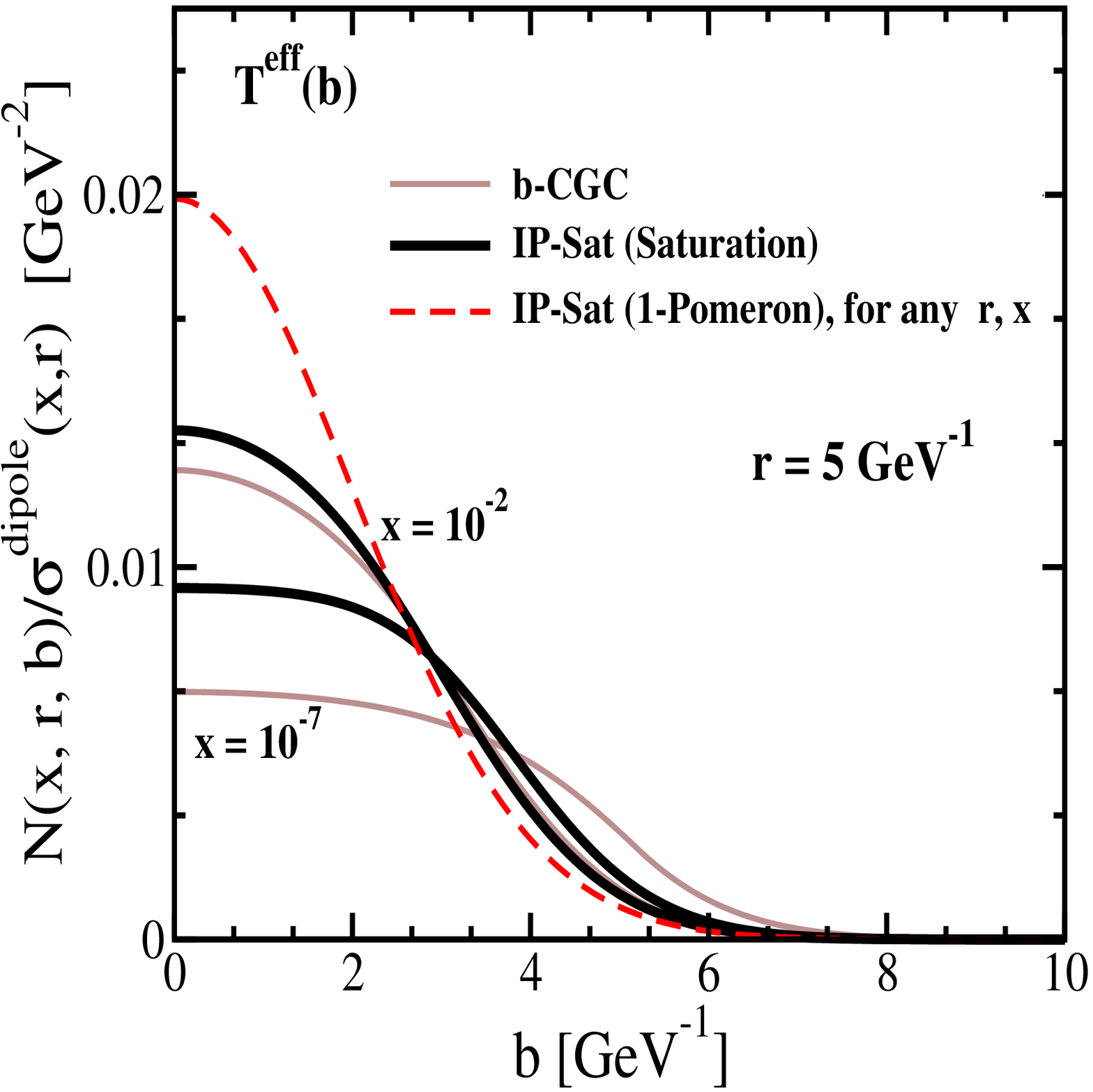}
\caption{Effective impact-parameter profile of the dipole amplitude defined as $T^{eff}(b)=\mathcal{N}(x,r,b)/\sigma^{dipole}(x,r)$ in different models as a function of impact-parameter $b$, for a fixed value of $x$ and two different values of  dipole transverse size $r$ (left panel), and for a fixed $r$ and two different values of $x$ (right panel). }
  \label{f-w6}
\end{figure}

%%%
Admittedly, the impact parameter dependence in saturation models lies in the domain of non-perturbative physics as commented previously and is, at present, put by hand and adjusted to data. A Gaussian profile is usually considered, but one could also try another profile whose Fourier transform leads to dips in the diffractive distribution. Therefore, the presence of dips cannot be considered, per se, as a signal of saturation.
But it is important to note that the main difference between a dipole model with linear and non-linear evolution (incorporating saturation effects through some specific model as those employed in this work) is that the former does not lead to the black-disc limit and, therefore, the dips do not systematically shift toward lower $|t|$ by increasing $W_{\gamma p}$, $1/x$, and $r$ or $1/Q$, while the latter does. 
%%%%%%%
Non-linear evolution evolves any realistic profile in $b$, like a Gaussian or Woods-Saxon distribution, and makes it closer to a step-like function in the $b$-space by allowing an increase in the periphery of the hadron (the dilute region) while limiting the growth in the denser center, see Fig. \ref{f-w6} for illustration.  This leads to the appearance of dips with non-linear evolution even if the dips were not present at the initial condition at low energies or for large $x$ (e.g. a Gaussian profile), or to the receding of dips towards lower values of $|t|$ even if they were already present in the initial condition (e.g. with a Woods-Saxon type profile). In \fig{f-w6}, we show the evolution of the effective impact-parameter profile of the dipole amplitude defined as $T^{eff}(b)=\mathcal{N}(x,r,b)/\sigma^{dipole}(x,r)$ with $x$ and $r$ in different models. It is clearly seen that in the saturation models, by increasing $1/x$ or $r$, the effective impact-parameter profile $T^{eff}(b)$ naturally evolves towards a step-like function with a dynamical median extended to a larger $b$ while, in contrast, in the 1-Pomeron dipole model $T^{eff}(b)$ does not change with $r$ and $x$.  Note that the typical impact-parameter of collisions is approximately related to the inverse of $|t|$, namely  $|t|\propto 1/b$, see Eqs.\,(\ref{am-i}), (\ref{vm}) or \eq{d-f}. Now, at small $|t|$, the typical collisions are mostly peripheral and  the system is in the dilute regime with a Gaussian profile. Therefore, saturation effects become less relevant, and there will be no dip in the $t$-distribution.  On the other hand, at large $|t|$ the typical collisions are central, and interactions probe the high-density region of the target proton. Then saturation effects become important and distort the impact-parameter profile leading to diffractive dips. 

%%%%%%%
The position of the dip in the $t$-distribution is presently rather model dependent. This is mainly due to the fact that the appearance of dips probes the dipole scattering amplitude in the saturation regime, where current available data at small $x$ do not constrain sufficiently the dipole models \cite{b-cgc-a,Albacete:2014fwa}.  The exact position of the dip can only be numerically computed  and depends on the effective dipole transverse size probed by the system (via a convolution between vector meson overlap wavefunction and the dipole amplitude) and impact-parameter profile of the saturation scale. Nevertheless, it is qualitatively  expected that the dip becomes stronger or moves to lower $|t|$ for the case that the saturation or unitarity effects probed by the system at a given kinematics and impact parameter become more important. The saturation scale in the IP-Sat and b-CGC models is approximately similar at the HERA kinematics for the typical impact-parameter probed in the total $\gamma^\star p$ cross-section of about $b\approx 2\div 3\,\text{GeV}^{-1}$ \cite{b-cgc-a}. However, at very small $x$ and large $|t|$ the effective impact-parameter profile of the dipole amplitude in these two saturation models is different. This is shown in \fig{f-w6} where it is seen that in the b-CGC model because of non-trivial correlations between $x$ and $b$, the effective impact-parameter profile of dipole tends to flatten sooner with lowering $x$ and/or increasing dipole transverse size $r$ compared to the IP-Sat model. Therefore,  the black disk limit is probed slightly faster in the b-CGC model than in the  IP-Sat model, and consequently the dip (or dips) appears at lower $|t|$ in the b-CGC model compared to the IP-Sat model. This general expectation is, remarkably, in accordance  with the results obtained from full computation for different vector mesons shown in \fig{f-vt2}. We also numerically verified that changing kinematics ($W_{\gamma p}$ and $Q$) does not alter this feature.

\section{conclusion}

In this paper, we investigated the exclusive production of vector mesons in high-energy collisions. We extended previous studies \cite{ip-sat0,ip-sat1} by using saturation models fitted to the most recent inclusive DIS data  and  including in our analysis the recent experimental diffractive data from the LHCb \cite{lhcb,lhcbn} and ALICE \cite{TheALICE:2014dwa} collaborations, and the combined HERA analysis \cite{Aaron:2009aa,Abramowicz:1900rp}.  We showed that the recent LHC data on diffractive $J/\psi$ photoproduction are in good agreement with the saturation/CGC predictions while there are some tensions between recent LHCb and ALICE data with the 1-Pomeron model and pQCD results, see Figs.\,\ref{f-vw1},\,\ref{f-vw2} (right panel). This can be considered as the first hint of saturation effects at work in diffractive photoproduction of vector mesons off proton at the LHC. 

We provided predictions for the total cross-section of diffractive photoproduction of  $J/\psi, \psi(2s)$ and $\rho$ within the gluon saturation/CGC picture at the LHC and future colliders. To single out the non-linear effects due to saturation, we also compared with those results obtained in the 1-Pomeron model. We also provided predictions for the ratio of diffractive production of $\psi(2s)$ to $J/\psi$, namely $R=\psi(2s)/J/\psi$ at HERA and the LHC.
We showed that while at high virtualities $R$ has little $|t|$ and $W_{\gamma p}$  dependence, it moderately increases with virtuality $Q$ at a fixed $W_{\gamma p}$.  We also found that the photoproduction ratio $R (Q=0)$ increases with $W_{\gamma p}$ and becomes  sensitive to different saturation models.

We showed that the $t$-differential cross-section of exclusive production of vector mesons in high-energy collisions offers a unique opportunity to probe the saturation regime. We quantified some theoretical uncertainties and showed that the appearance of a dip or dips  in the diffractive $t$-distribution of the different  vector mesons ($J/\psi, \psi(2s)$, $\phi,\rho$) is a robust prediction of the saturation picture.  In non-saturation models, dips are either absent or expected to lie at larger $|t|$ and not to shift towards smaller $|t|$ with increasing energy. The position of the dip is presently rather model dependent, see \fig{f-vt0}. This is mainly due to the fact that the appearance of dips probes  the dipole scattering amplitude at large dipole sizes in the saturation regime, where current available data at small $x$ do not constrain sufficiently the dipole models \cite{b-cgc-a,Albacete:2014fwa}.
On the positive side, future experimental data, in particular at the LHeC, on the energy and $t$ diffractive distributions of different  vector meson production off protons, will provide valuable constraints on saturation models and allow us to unravel the relevance of non-linear effects in the accessible kinematic region.

We recall that the $t$-distribution of all vector mesons, as well as DVCS, at HERA, can be correctly reproduced by fixing the impact-parameter profile of the colour dipole amplitude at small $|t|$, despite the fact that the vector meson and DVCS wave functions are very different \cite{ip-sat-a,b-cgc-a}. This strongly hints at universality of the extracted impact-parameter distribution of gluons in the periphery of the proton.  On the other hand, at large $|t|$ where we do not have currently experimental data, one can probe the transverse spatial distribution of gluons in the center of the proton where the black-disc limit could be at work. Therefore, the $t$-distribution of diffractive vector mesons would provide the most important information on the relevance of saturation dynamics. Besides, the  impact parameter distribution of gluons  in protons and nuclei (a natural extension of our work that can be explored in electron-nucleus colliders \cite{eic,lhec}, see also \cite{Frankfurt:2011cs,Caldwell:2010zza,ip-g1,tu}) is a crucial ingredient for a detailed characterisation of the initial conditions in heavy ion collisions.  Note that the effects of fluctuations and correlations on the proton are not incorporated into our formulation. This is an important issue that certainly deserves separate study.

Note that diffractive vector meson production off a nucleus is quite different from a proton target. The diffractive interaction with the nuclear target can either be elastic (coherent) or inelastic (incoherent) - in the latter case the nucleus subsequently radiates a photon or breaks up into colour neutral fragments, while in the former, the nucleus stays intact.  The coherent cross section is obtained by averaging the amplitude before squaring it, $|\langle\mathcal{A}\rangle_ N |^2$, and the incoherent one is the variance of the amplitude with respect to the initial nucleon configurations N of the nucleus $\langle |\mathcal{A}|^2\rangle_ N -|\langle  \mathcal{A}\rangle_ N |^2$  which according to the Good-Walker picture measures the fluctuations or lumpiness of the gluon density inside the nucleus. In the case of a nucleus, the diffractive production rate is controlled by two different scales of $1/R_p$ and $1/R_A$ with $R_p$ and $R_A$ being the proton and nucleus size.  At momentum scales corresponding to the nucleon size $|t|\sim 1/R_p^2$ the diffractive cross section is almost purely incoherent. The $t$-distribution in coherent diffractive production off nucleus gives rise to a dip-type structure for both saturation and non-saturation models, while in the case of incoherent production at small $|t|$, both saturation and non-saturation models do not lead to dips \cite{ip-g1,tu}. This is in drastic contrast to the diffractive production off proton where only saturation models lead to dip-type structure in the $t$-distribution at values of $|t|$ that can be experimentally accessible. Therefore, diffractive production off nucleus is a sensible probe of unitarity effects at the nuclear level while being less sensitive to the unitarity limit and saturation effects inside the proton.

Finally, note that diffractive dips in $t$-distribution were also observed in elastic hadronic reactions \cite{dip-el,dip-el-2}. However, it remains to be understood whether the origin of the dips in elastic hadronic reactions and diffractive DIS is the same. In contrast to diffractive DIS,  the differential cross-section of elastic proton-proton collisions is not currently computable in the weak coupling regime due to the absence of a large scale, and some phenomenological models are often employed (for a review see Ref.\,\cite{rev-el}). Nevertheless, in both cases multiple parton interactions or multiple Pomeron exchanges seem to play an important role in the appearance of dips in the $t$-distribution, see e.g. \cite{all-el,Brogueira:2011jb}.

\begin{acknowledgments}
We thank Jorge Dias de Deus, Edmond Iancu, Paul Laycock, Genya Levin, Magno Machado and Jan Nemcik for useful discussions. The work of NA is supported by the European Research Council grant HotLHC ERC-2011-StG-279579; by the People Programme (Marie
Curie Actions) of the European Union Seventh Framework Programme FP7/2007-2013/
under REA grant agreement n318921; by Ministerio de Ciencia e Innovaci\'on of Spain under project FPA2011-22776; by Xunta de Galicia (Conseller\'{\i}a de Educaci\'on and Conseller\'\i a de Innovaci\'on e Industria - Programa Incite); and by the Spanish Consolider-Ingenio 2010 Programme CPAN and FEDER. 
The work of AHR is supported in part by Fondecyt grant 1110781.  
\end{acknowledgments}

%\newpage

\newpage

 \end{document}